\documentclass[usenatbib,useAMS english]{mnras}
\usepackage[fleqn]{amsmath}
\usepackage{array,multirow,graphicx}
\usepackage{txfonts}
\usepackage{lmodern}
\bibpunct{(}{)}{;}{a}{}{,}
\usepackage{floatrow}
\usepackage{array}
\usepackage{relsize}
\usepackage{bm}
\usepackage{listings}
\usepackage{url}
\usepackage[font=small,labelfont=bf,skip=0pt]{caption}

\graphicspath{{./figs/}}

\newcommand{\zo}{$z\!=\!0$}
\newcommand{\HI}{H\,{\sc i}}
\newcommand{\Htwo}{H$_{2}$ }
\newcommand{\DS}{{\sc Dark Sage}}

\newfloatcommand{capbtabbox}{table}[][\FBwidth]

\def\app#1#2{%
\mathrel{%
\setbox0=\hbox{$#1\sim$}%
\setbox2=\hbox{%
\rlap{\hbox{$#1\propto$}}%
\lower1.1\ht0\box0%
}%
\raise0.25\ht2\box2%
}%
}
\def\approxprop{\mathpalette\app\relax}

\title[Dissecting galaxies' angular momenta \& neutral gas]{Connecting and dissecting galaxies' angular momenta and neutral gas in a hierarchical universe: cue {\sc Dark Sage}}

\author[A.~R.~H.~Stevens et al.]{Adam R.~H.~Stevens,$^1$\thanks{E-mail: adam.stevens@uwa.edu.au} Claudia del P.~Lagos,$^{1,2}$ Danail Obreschkow$^1$ \newauthor and Manodeep Sinha$^{2,3}$ \\
$^1$International Centre for Radio Astronomy Research, The University of Western Australia, Crawley, WA 6009, Australia\\
$^2$Australian Research Council Centre of Excellence for All Sky Astrophysics in 3 Dimensions (ASTRO 3D)\\
$^3$Centre for Astrophysics and Supercomputing, Swinburne University of Technology, Hawthorn, VIC 3122, Australia}

\newcommand{\appropto}{\mathrel{\vcenter{
  \offinterlineskip\halign{\hfil$##$\cr
    \propto\cr\noalign{\kern2pt}\sim\cr\noalign{\kern-2pt}}}}}

\begin{document}

\defcitealias{ob16}{O16}
\defcitealias{og14}{OG14}
\defcitealias{stevens16}{SCM16}
\defcitealias{sb17}{SB17}

\pagerange{\pageref{firstpage}--\pageref{lastpage}} \pubyear{2018}

\maketitle

\label{firstpage}

\begin{abstract}
We explore the connection between the atomic gas fraction, $f_{\rm atm}$, and `global disc stability' parameter, $q$, of galaxies within a fully cosmological context by examining galaxies in the \DS~semi-analytic model.  The $q$ parameter is determined by the ratio of disc specific angular momentum to mass, i.e.~$q \! \propto \! j_{\rm disc}/m_{\rm disc}$.  \DS~is well suited to our study, as it includes the numerical evolution of one-dimensional disc structure, making both $j_{\rm disc}$ and $q$ \emph{predicted} quantities.  We show that \DS~produces a clear correlation between gas fraction and $j_{\rm disc}$ at fixed disc mass, in line with recent results from observations and hydrodynamic simulations. This translates to a tight $q$--$f_{\rm atm}$ sequence for star-forming central galaxies, which closely tracks the analytic prediction of \citeauthor{ob16} The scatter in this sequence is driven by the probability distribution function of mass as a function of $j$ (PDF of $j$) within discs, specifically where it peaks.  We find that halo mass is primarily responsible for the peak location of the PDF of $j$, at least for low values of $q$.  Two main mechanisms of equal significance are then identified for disconnecting $f_{\rm atm}$ from $q$.  Mergers in the model can trigger quasar winds, with the potential to blow out most of the gas disc, while leaving the stellar disc relatively unharmed.  Ram-pressure stripping of satellite galaxies has a similar effect, where $f_{\rm atm}$ can drop drastically with only a minimal effect to $q$.  We highlight challenges associated with following these predictions up with observations.

\end{abstract}


\section{Introduction}
\label{sec:intro}

The most in-depth physical model for any system should be described explicitly in terms of fundamental properties, backed by a theory for how those properties evolve with time.  In the case of galaxies, the two most obvious fundamental quantities are mass (energy), $m$, and specific angular momentum, $j$.  Following the pioneering paper of \citet{fall83}, significant efforts have been made in previous decades to model isolated galaxies explicitly in terms of their angular momentum \citep*{dalcanton97,firmani00,bosch01,dutton09}.  More recently, there have been advancements in the treatment and evolution of galaxies' angular momenta in semi-analytic models (\citealt{stringer07}; \citealt{guo11}; \citealt*{stevens16}, hereafter SCM16).  These works not only ground galaxies in a fully cosmological context, but also root their evolution as a whole on specific angular momentum.  With the complement of a wide variety of studies of galaxies' specific angular momenta both in observations \citep*{rf12,og14,cortese16,swinbank17,okamura18,posti18,sweet18} and hydrodynamic simulations \citep{genel15,pedrosa15,teklu15,lagos17,lagos18,stevens17,el18,schulze18}, the movement to describe galaxies in terms of fundamental properties has been \emph{gaining momentum}.

Several recent works have highlighted that the specific angular momenta of galaxies are tied to their fraction of cold baryons in a gaseous state \citep{ob15,ob16,lagos17,zoldan18}.  Gas fractions tell us about the self-regulatory nature of galaxy growth; as galaxies increase in stellar mass, their gas fractions tend to decrease \citep[e.g.][]{haynes84,catinella10,catinella18} but remain closely tied to their specific star formation rates \citep{brown15}, meaning their \emph{relative} (i.e.~fractional) in-situ growth rates decline.  Understanding processes or properties that regulate gas fractions is critical to furthering our narrative for how galaxies evolve.

As analytically shown by \citet[][hereafter O16]{ob16}, the gas fractions of idealised disc galaxies can be interpreted as a natural result of angular-momentum regulation.  Those authors argued that the fraction of a galactic disc (composed initially of an atomic `warm neutral medium') that is Toomre stable \citep[i.e.~with $Q\!>\!1$ -- cf.][]{toomre64,binney87} will remain atomic, while the rest should collapse to become molecular and/or form stars in relatively short time-scales.  For cold baryonic discs with exponential surface density profiles and simple rotation curves (either constant or asymptotic), \citetalias{ob16} derived that this fraction is set by the disc's $j/m$ ratio.  While this model has proven highly informative, it is limited in the sense that it does not account for variations in disc structure, nor the hierarchical assembly and evolutionary physics of galaxies.  We are therefore motivated to explore the idea that angular momentum regulates galaxies' cold-gas content in a more sophisticated model of galaxy evolution that is embedded in the standard $\Lambda$CDM cosmological model.

To fully explore the gas fraction--angular momentum connection requires cosmological simulations.  Either, one can investigate galaxies in fully hydrodynamic simulations, or one can use a semi-analytic model, which evolves galaxies in an $N$-body simulation as a post-processing step \citep[for a review, see][]{somerville15}.  To make predictions for the global stability or specific angular momentum of a galaxy requires self-consistent evolution of the galaxy's structure.  Hydrodynamic simulations naturally excel at this, but semi-analytic models also come with several advantages.  For one, galaxies in semi-analytic models are typically very well calibrated to observed global statistics, and thus more accurately represent galaxies in the real Universe in terms of their integrated properties.  This comes as a result of the relative computational efficiency of these models, which allows for a fast exploration of the `free' parameter space.  Moreover, larger simulated volumes can be explored, leading to a wider range of environments being probed.  Another favourable aspect is that galaxy components such as discs and bulges are defined \emph{a priori} in semi-analytic models, such that any debate on how these should otherwise best be decomposed (which would inevitably introduce uncertainty) is circumvented.  While classical semi-analytic models tend to \emph{prescribe} disc structure, modern models are beginning to make a concerted effort to numerically \emph{evolve} the one-dimensional structure of discs (\citealt{stringer07}; \citealt{fu10}; \citetalias{stevens16}) and hence calculate galaxy evolution processes pseudo-locally within annuli, rather than on purely global scales.  

In this paper, we study galaxies within the \DS~semi-analytic model \citepalias{stevens16} to further develop the picture of how galaxies' cold-gas content and angular momenta are connected, and what can cause these properties to become dissociated. The main feature of \DS~is that it numerically evolves the angular-momentum structure (in tandem with the radial structure) of galaxy discs.  \DS~is particularly ideal for our study, as it has already been shown to predict (or \emph{post}dict) the right specific angular momenta of stellar discs for all masses $>\!10^9\,{\rm M}_{\odot}$ \citepalias{stevens16}, as well as reproduce the observed \HI-to-stellar mass fraction of galaxies for the same mass range \citep[][hereafter SB17]{sb17}, thereby providing us with a solid platform to work from.

Our paper is structured as follows.  In Section \ref{sec:model}, we briefly describe the main features of \DS, for which new features are elaborated in Appendix \ref{app:updates}.  In Section \ref{sec:mj}, we highlight the connections between the mass, specific angular momentum, and gas fraction of \DS~galaxies.  We then extensively analyse these galaxy properties at various redshifts in Section \ref{sec:causal}, identifying where the connections between them come from, and how this compares to the analytic model of \citetalias{ob16}.  We go on to examine situations where the otherwise tight relation between cold-gas content and the `global disc stability' parameter (given by the ratio of specific angular momentum to mass) breaks down, specifically from mergers and ram-pressure stripping, in Section \ref{sec:poor}.  Our results and their context are summarised in Section \ref{sec:summary}.


\section{Overview of D{\small ark} S{\small age} and its updates}
\label{sec:model}
\DS~is a modern semi-analytic model of galaxy formation, developed by \citetalias{stevens16}, that extends the well-established, classical {\sc sage}\footnote{Semi-Analytic Galaxy Evolution} model \citep{croton16} to numerically evolve (rather than prescribe) the one-dimensional structure of galactic discs.  Instead of treating the gas and stellar components of galactic discs as singular baryonic reservoirs, each is broken into 30 annuli, whose edges are fixed in specific angular momentum (and thus are radially adaptive).  

Each annulus has its own H$_2$/\HI~fraction, forms its own stars, is subjected to different feedback strengths from stars and the central supermassive black hole, and is stripped by its environment and impacted by mergers differently.  While independent in the calculation of these specific processes, integrated halo and galaxy properties also factor in to each galaxy's overall evolution, and so the annuli evolve together coherently to form quasi-continuous discs.  Mass can also flow directly between annuli when an annulus becomes gravitationally unstable.  For full details, see \citetalias{stevens16}.

Already, \DS~has been used to show how disc instabilities regulate the relationship between the mass and specific angular momentum of stellar discs \citepalias{stevens16}; how ram-pressure stripping of the interstellar medium is responsible for the observed variation in \HI~content of galaxies as a function of environment, whereas hot-gas stripping causes quenching of satellites \citepalias{sb17}; and that the most \HI-rich galaxies in the local Universe most likely are a natural consequence of evolving in haloes with high spins \citep{lutz18}.

The model codebase is entirely publicly available,\footnote{\label{foot:github}\url{https://github.com/arhstevens/DarkSage}} and catalogues from the \citetalias{stevens16} version of the model, including light cones and magnitudes, are constructible and downloadable through the Theoretical Astrophysical Observatory\footnote{\url{https://tao.asvo.org.au/}} \citep{tao}.  \DS~is under constant development; as a checkpoint, in Appendix \ref{app:updates}, we describe updates since the version presented in \citetalias{sb17}.  These include (i) an update to how gas is distributed during cooling (following the results of \citealt{stevens17} on the profiles of recently cooled gas in the EAGLE hydrodynamic simulations), (ii) a more thorough interpretation of disc size for prescriptions that require a size measure as an input, and (iii) the inclusion of dispersion support in the centres of discs.  Outside of these updates, we use the same default features as in \citetalias{sb17}.  For example, we maintain the metallicity-based prescription for breaking hydrogen in the interstellar medium into its atomic and molecular components (\citealt{mckee10}, for which our implementation is based on that of \citealt{fu13}) for this work.  Where results pertain to specific aspects of the model, we discuss those aspects in more detail (i.e.~mergers and quasar winds in Section \ref{ssec:mergers}, and gas stripping in Section \ref{ssec:rps}).  The reader is \emph{strongly} encouraged to see \citetalias{stevens16} for further details on the model's design.  

The model has been run on the standard merger trees of the Millennium simulation \citep{millennium}.  To be consistent with the \emph{WMAP-1} cosmology used for the simulation \citep{wmap1}, we assume $h\!=\!0.73$ for all results in this paper.  To match the assumptions of \DS, where necessary, we also modify compared observational data to be consistent with a \citet{chabrier03} stellar initial mass function: we assume $0.66\, m_{\rm *,Chabrier} \! = \! 0.61\, m_{\rm *,Kroupa}$ \citep{madau14}.

\subsection{Recalibration of the model}
\label{ssec:recalibration}
Whenever prescriptions are updated or added to a semi-analytic model, the free parameters of the model require recalibration.  We considered a variety of observational data at $z\!=\!0$ when calibrating \DS~after applying the updates in Appendix \ref{app:updates}.  Calibration was performed manually by visual comparison against observational data (followed by $\chi^2$ measurements where relevant).  We endeavoured to simultaneously reproduce (i) the total stellar mass function \citep*{baldry08}, as well as the contributions from bulge- and disc-dominated galaxies \citep{moffett16}; (ii) the mean \HI-to-stellar mass fraction as a function of stellar mass \citep{brown15}; (iii) the \HI~mass function \citep{zwaan05,martin10}; (iv) the \Htwo mass function \citep*{keres03}; (v) the black hole--bulge mass relation \citep*{scott13}; (vi) the Baryonic Tully--Fisher relation for gas-dominated galaxies \citep*{stark09}; (vii) the stellar mass--gas metallicity relation \citep{tremonti04}; and (viii) the \HI~and \Htwo surface density profiles for analogues of galaxies presented by \citet{leroy08}.  Most of these were adopted from the calibration procedures of \citet{croton16} and \citetalias{stevens16}.  Each of these constraints was met with varying degrees of accuracy.  Our philosophy was to cover a broad range of galaxy properties and statistics by altering as few parameters as little as possible, rather than fine-tuning parameters to match a few of these with absolute quantitative precision.%
\footnote{We have not differentiated between `primary' and `secondary' constraints for this incarnation of the model.}  
Overall, the model is largely representative of these datasets.  Calibration figures are presented in Appendix \ref{app:constraints}, where other, recent sources of observational data are also compared.

\begin{table*}
	\centering
	\begin{tabular}{l l l l l} \hline
		Parameter & Description & Value & \citetalias{sb17} value & \citetalias{stevens16} value\\ \hline
		$\epsilon_{\rm SF}$ & Passive star formation efficiency from H$_2$ [$10^{-10}$ yr$^{-1}$] & 1.3 & 1.3 & 3.96 \\
		$Y$ & Yield of metals from new stars & 0.03 & 0.025 & 0.025 \\
		$\Sigma_{0,\rm gas}$ & Surface density scaling for supernova reheating [$\mathrm{M}_{\odot}\, \mathrm{pc}^{-2}$] & 2.0 & 8.0 & 8.0 \\ 
		$f_{\rm move}$ & Fraction of unstable gas that moves to adjacent annuli & 0.5 & 0.3 & 0.3 \\
		$f_{\rm BH}$ & Rate of black hole growth during quasar mode accretion & 0.1 & 0.03 & 0.03 \\ 
		$\kappa_{\rm R}$ & Radio mode feedback efficiency & 0.02 & 0.035 & 0.035 \\
		$\phi$ & Clumping factor normalisation for H$_2$ fraction & 1.0 & 3.0 & N/A \\
		$w_{90}$ & Weighting of $r_{*,90}$ in calculating stellar disc scale radii & 2.0 & N/A & N/A \\ \hline
	\end{tabular} 
	\caption{Free parameters in \DS~that we have updated for this paper, as necessary to accommodate the new model features.  Compared are the values of those parameters in previously published versions of the model (\citealt{sb17} and \citealt{stevens16}, respectively).  All other parameters remain unchanged to \citetalias{stevens16} (see their table 1).  The new parameter, $w_{90}$, is described further in Appendix \ref{ssec:rs}.}
	\label{tab:params}
\end{table*}

We also explored the stellar surface density profiles for the \citet{leroy08} galaxy proxies, as well as the H$_2$-to-\HI~ratio of galaxies as a function of stellar mass as compared to \citet{catinella18}.  The parameter set we settled on unfortunately resulted in too-steep stellar profiles and too-high molecular fractions relative to those respective datasets, on average.  Of course, there is uncertainty in how the model data are sliced to compare to observations, as are there significant systematic uncertainties in the observations.  So these are not necessarily `failures' of the model per se.  Future investigation of these aspects of \DS~galaxies is nevertheless warranted.

Table \ref{tab:params} compiles the values for calibrated \DS~parameters, listing only those that have changed since \citetalias{stevens16}, and compares to their original values.  For the mathematical context of the parameters otherwise not discussed in this paper, the reader is referred to \citetalias{stevens16} and \citetalias{sb17}.  We note that while the yield of new metals generated from star formation was treated as a free parameter, the new value of 0.03 is very close to what one should derive theoretically for a \citet{chabrier03} initial mass function (which \DS~already assumes for the recycling fraction and observational constraints) with a \citet*{conroy09} model for simple stellar populations; the precise yield one gets from that is 0.02908.


\section{Relating galaxies' masses and specific angular momenta with their gas fractions}
\label{sec:mj}

Recent works have pointed out that perhaps the most important third variable for relating the mass and angular momentum of galaxies is gas fraction \citep[e.g.][]{ob15}.  This is true even when mass and specific angular momentum are calculated using only stellar content; this has been shown, for example, with the EAGLE\footnote{Evolution and Assembly of GaLaxies and their Environments \citep{schaye15}.} simulations \citep{lagos17}, complemented by results from the {\sc gaea} semi-analytic model \citep{zoldan18}.  One simple, physical explanation for this is that at fixed stellar mass, a galaxy with less angular momentum will be relatively compact and dense, and hence should have had an easier time forming stars from its available gas over its recent history.  That galaxy would, therefore, have a lower gas fraction at present (and in the past).  Equivalently though, the galaxy with more angular momentum could have had a recent period of gas accretion, with that gas carrying a higher $j$ than previously exisiting gas in the galaxy; a result of cold dark matter is that material accreted at later times is expected to carry higher $j$ \citep{catelan96}.

\begin{figure}
\centering
\includegraphics[width=\textwidth]{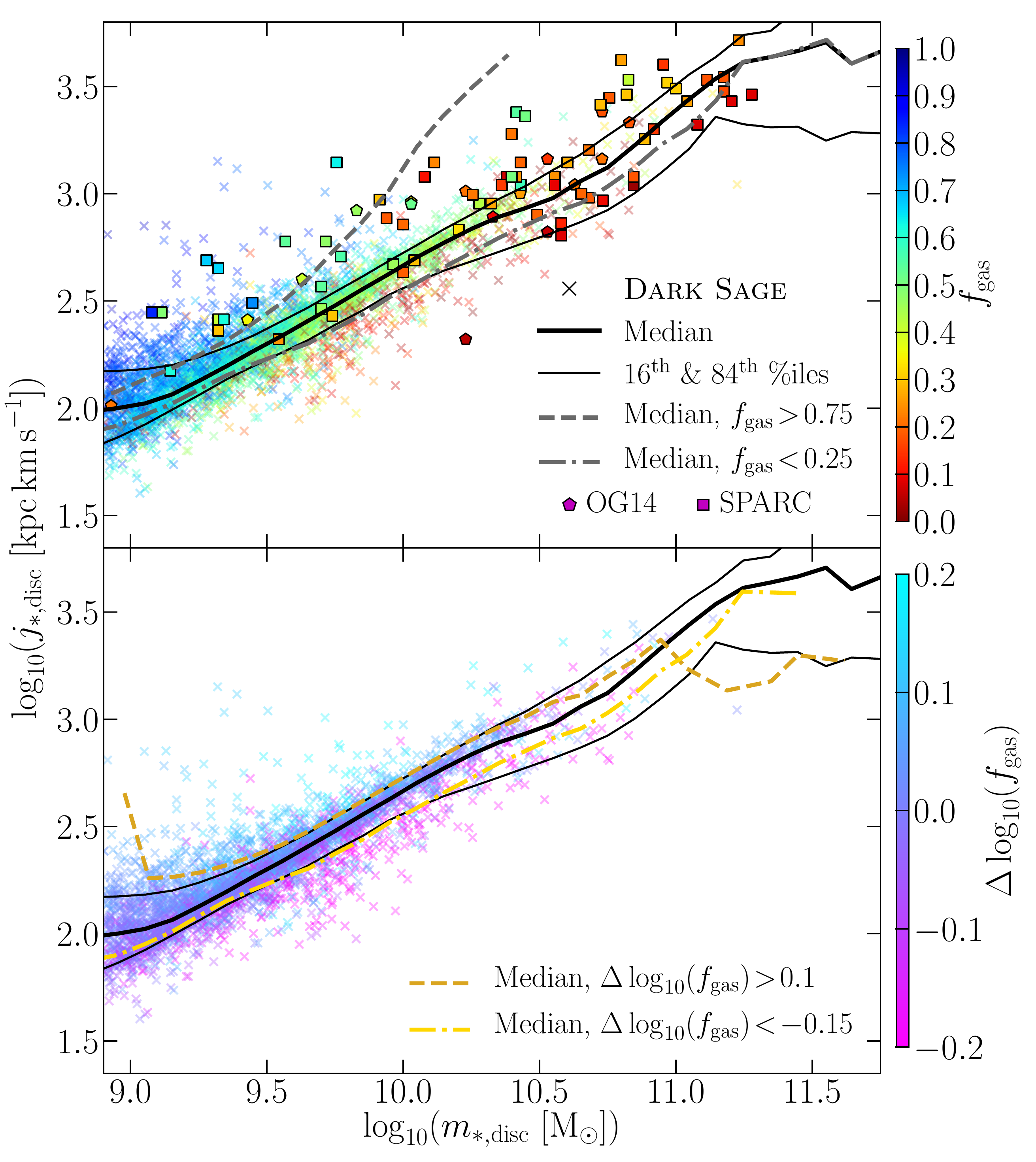}
\caption{Specific angular momentum as a function of mass for stellar discs, coloured by the discs' gas fractions.  Pseudo-bulges are considered to be part of the discs (whereas classical bulges, either grown from mergers or instabilities, do not contribute to this plot).  Only central galaxies from \DS~at $z\!=\!0$ with $m_{\rm *,disc} \! > \! 0.5\,m_*$ in haloes of at least 100 particles are included; the left-hand edge of the $x$-axis was chosen as the median stellar disc mass for those exactly at the 100-particle limit.  Points for 4000 representative galaxies within the axes are shown.  The median trend and scatter is tracked by the solid curves, where the dashed and dot-dashed track the median trends for gas-rich and gas-poor galaxies, respectively.  These were built from bins of minimum width 0.1\,dex with at least 20 galaxies per bin.  The top panel colours by absolute gas fraction on a linear scale.  The bottom panel colours by the logarithm of the fractional difference in gas fraction from the median value of galaxies at the same stellar disc mass.  Compared in the top panel are observational data of spiral galaxies from \citet{og14} and SPARC \citep{lelli16,posti18}: see text in Section \ref{sec:mj} for details.}
\label{fig:jm_gasfrac}
\end{figure}

Indeed, the galaxies in \DS~also highlight the same correlation between disc gas fraction and stellar specific angular momentum for fixed stellar mass, as we show in Fig.~\ref{fig:jm_gasfrac}.  Here, we have defined gas fraction as 
\begin{equation}
f_{\rm gas} \equiv \frac{m_{\rm H\,{\LARGE{\textsc i}}+H_2}}{X\,m_{\rm disc}}~,
\end{equation}
where
\begin{equation}
m_{\rm disc} \equiv m_{\rm *,disc} + X^{-1} m_{\rm H\,{\LARGE{\textsc i}}+H_2}\,,
\label{eq:mdisc}
\end{equation}
and $X$ is the fraction of gas in the form of hydrogen [\DS~assumes $X\!=\!0.75(1-Z)$, where $Z$ is the gas disc's mean metallicity].  Compared in Fig.~\ref{fig:jm_gasfrac} are the 16 observed, local spiral galaxies of \citet[][hereafter OG14]{og14}, and the SPARC\footnote{\emph{Spitzer} Photometry and Accurate Rotation Curves} galaxies studied in \citet{posti18}. The angular momenta for all these galaxies have been measured precisely using spatially resolved velocity and surface brightness data.  While we only account for stellar \emph{disc} content for \DS~galaxies in Fig.~\ref{fig:jm_gasfrac}, \emph{all} stellar mass contributes to both axes for the \citetalias{og14} data.  This is because the \citetalias{og14} galaxies with significant bulge-to-total ratios (the highest of which is quoted to be 0.32) are thought to host primarily pseudo-bulges, rather than classical%
\footnote{`Classical' in this case means high S\`{e}rsic index and typically supported by dispersion.} 
bulges \citep[see][]{sweet18}. The comparison is thus fairer, as pseudo-bulges in \DS~are inherently considered to be part of the disc.  For SPARC, we have used the disc-only measurements for stellar mass and $j$ from \citet{posti18}.  For galaxies of Hubble-type Sbc and later, the disc-only measurements are the same as the total measurements; any bulge-like component was assumed be to a pseudo-bulge that was counted as part of the disc \citep{lelli16}. \HI~masses for SPARC galaxies were obtained and cross-referenced from those published in \citet{lelli16}.  Note that we have deviated from the plotting convention in \citetalias{stevens16}, where a rudimentary decomposition of pseudo-bulges and discs was made in post-processing.  While the sample is humble in number, the observations exhibit the same trends as the result of \DS.

\begin{figure}
\centering
\includegraphics[width=\textwidth]{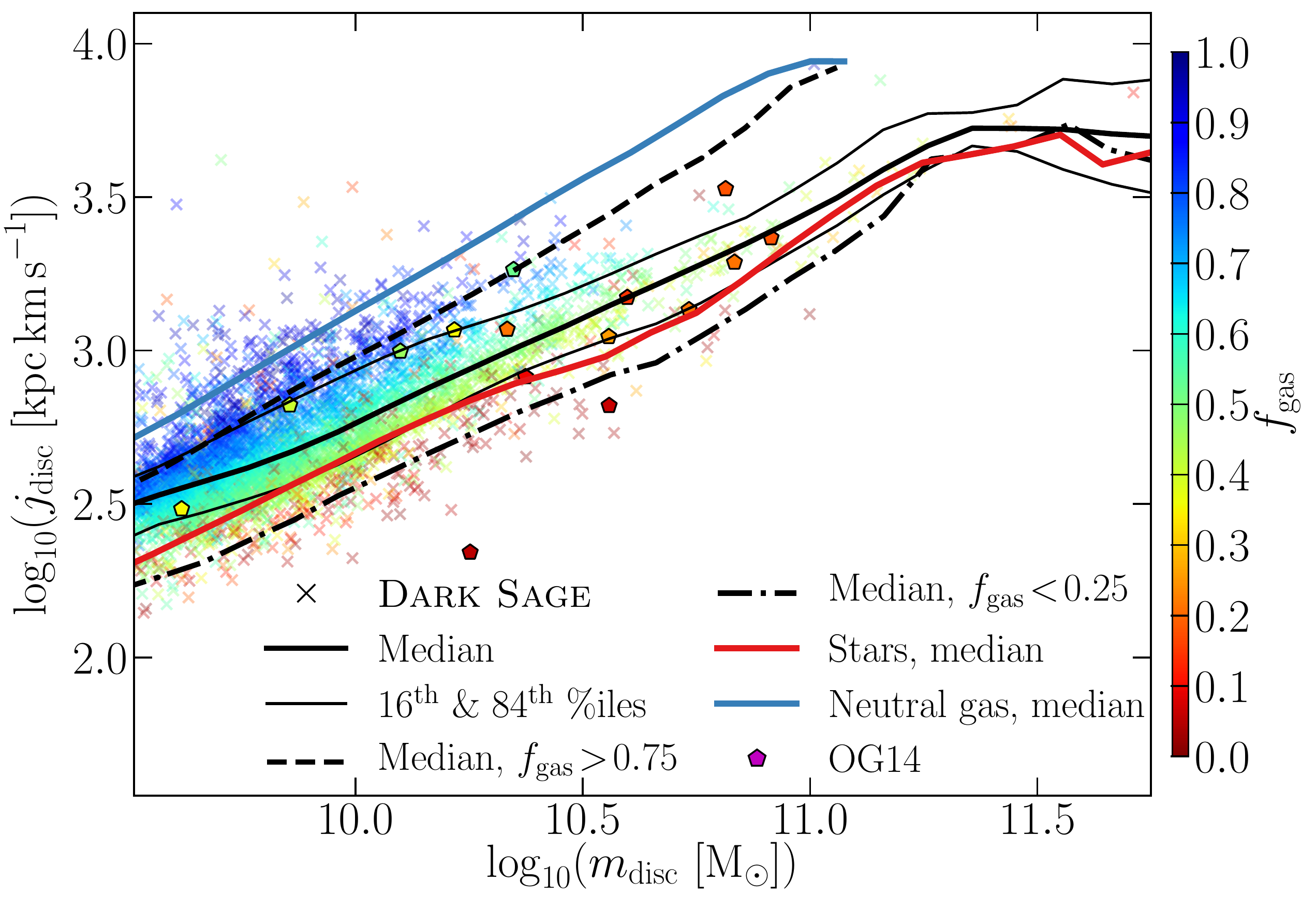}
\caption{Specific angular momentum as a function of mass for `all' disc content in galaxies, i.e. that from stars and neutral gas ($m_{\rm H\,{\LARGE{\textsc i}}+H_2} / X$).  This follows the same plotting conventions and galaxy cuts as Fig.~\ref{fig:jm_gasfrac}.  Included are the median trends for discs when stellar content and neutral gas are considered separately.}
\label{fig:jm_gasfrac_bary}
\end{figure}

If one examines the \emph{total} specific angular momentum of a disc (i.e.~from stars \emph{and} neutral gas), the correlation with gas fraction at fixed mass becomes even clearer.  Comparing to the same observational data from \citetalias{og14}, we show this for \DS~galaxies in Fig.~\ref{fig:jm_gasfrac_bary}, where we define
\begin{equation}
j_{\rm disc} \equiv \frac{m_{\rm *,disc}\, j_{\rm *,disc} + X^{-1}\, m_{\rm H\,{\LARGE{\textsc i}}+H_2}\, j_{\rm H\,{\LARGE{\textsc i}}+H_2}}{m_{\rm disc}}~.
\label{eq:jdisc}
\end{equation}
Because the observational data we compare against only include \HI~and H$_2$ with a helium correction, we have excluded the ionized contribution of \DS~gas discs in both the definitions of mass (Equation \ref{eq:mdisc}) and specific angular momentum.

Over-plotted in Fig.~\ref{fig:jm_gasfrac_bary} are the mean trends for the separate disc components of stars (same as in Fig.~\ref{fig:jm_gasfrac}) and neutral gas.  Because stars are preferentially born of the low-$j$ gas in a disc, the stellar $m$--$j$ sequence lies $\gtrsim \! 0.5$\, dex below the gas $m$--$j$ sequence.  The gas fraction of a galaxy then effectively weights an average between these two sequences for the total-baryon $m$--$j$ sequence.  At low masses, galaxies typically have higher gas fractions, hence the gas sequence has a stronger weighting; whereas, at high masses, the total-baryon $m$--$j$ sequence tends towards the stellar sequence.  To rephrase this, all discs are born purely of gas, and their initial and subsequent evolution of angular momentum drives the amount of stars they will form.

Although the \citetalias{og14} data include both \HI~and H$_2$, the majority of cold-gas data in the literature are restricted to \HI.  While concerted efforts to survey galaxies in both \HI~and CO to gauge the total cold-gas content of galaxies have become significant in the last decade \citep{leroy09,saintonge11,saintonge17,catinella18}, the relative difficulty in measuring H$_2$ masses over \HI~means that \HI~data will continue to outpace H$_2$ for the foreseeable future.  Additionally, the \citetalias{ob16} model that we compare against in the following sections only applies for \HI, not H$_2$.  As such, the majority of results hereafter pertain to the \emph{atomic} gas fractions of galaxies.


\section{The causal connection: galaxies'  atomic gas fractions and the `global disc stability' parameter}
\label{sec:causal}

As shown by \citetalias{ob16} (and introduced in this paper in Section \ref{sec:intro}), if one assumes galaxy discs are exponential and that all Toomre-stable gas in the disc remains atomic, while the rest of the gas becomes either molecular or forms stars, the resulting atomic gas fraction of the disc is described entirely by a `global disc stability' parameter $q$:
\begin{subequations}
\label{eq:qfatm}
\begin{equation}
f_{\rm atm} \equiv \frac{m_{\rm H\,{\LARGE{\textsc i}}}}{X\, m_{\rm disc}} \simeq {\rm min}\left[1,~2.5\,q^{1.12}\right]\,,
\end{equation}
\begin{equation}
q \equiv \frac{\sigma_{\rm gas}\, j_{\rm disc}}{G\, m_{\rm disc}}\,,
\end{equation}
\end{subequations}
where $\sigma_{\rm gas}$ is the radial velocity dispersion of the warm neutral medium, assumed to be constant in both \citetalias{ob16} and \DS~(discussed further toward the end of this section).  In fact, \citetalias{ob16} present a more detailed expression for $f_{\rm atm}(q)$, which requires a numerical integral, and depends on the ratio of the exponential disc scale radius to the radius where the rotation curve asymptotes, denoted $a$ (see equations 4 and 11--13 of \citetalias{ob16}).  Equation (\ref{eq:qfatm}a) simply approximates this.  Because $q$ is directly proportional to the disc's specific angular momentum, this analytic relationship offers a theory behind \emph{how} the angular momenta and gas fractions of galaxy discs become so strongly related.  But this model is limited not only in its assumptions about disc structure, but also because it excludes any consideration of hierarchical structure formation and galaxy evolution physics.

With \DS, we can re-examine the $q$--$f_{\rm atm}$ relation with a more comprehensive consideration of relevant astrophysical processes (accretion, star formation, feedback, mergers, environment, et cetera).  We further allow freedom in the distribution of mass and rotation curves of discs -- both are evolved numerically using the disc annuli (i.e.~there are no strict assumptions of exponentiality).  \DS~also includes a breakdown of cold gas into its atomic and molecular components that is based on local density and metallicity (and thus is determined independently from the net specific angular momentum of the disc), whereas the \citetalias{ob16} model simply takes $f_{\rm atm}$ as the fraction of disc mass that is Toomre unstable, defined as $Q\!<\!1$ \citep[see][]{toomre64,binney87}.  

The local Toomre stability of discs plays a role in determining both $q$ and $f_{\rm atm}$ for \DS~discs, but in a different way to the \citetalias{ob16} model.  Whenever a disc annulus in \DS~becomes unstable (i.e.~it has more mass than it can support from its own self-gravity), it is brought back up to $Q\!=\!1$ through a combination of stellar-feedback outflows (from a triggered starburst) and transferral of mass to adjacent annuli (in proportion such that angular momentum is conserved).  Instabilities in \DS~tend to cascade into the central annulus, where excessive (unstable) stellar mass is then transferred to the bulge (further details can be found in section 3.7 of \citetalias{stevens16}).  This reduces $m_{\rm disc}$ and raises $j_{\rm disc}$ (as it is low-$j$ mass that is removed from the disc), thereby raising $q$.  It also raises $f_{\rm atm}$, as lower disc density means less of the gas will form molecules.  The demand that $Q \! \geq \! 1$ everywhere in a disc therefore leads to a positive correlation between $q$ and $f_{\rm atm}$.  In a sense, this is contrasting to the \citetalias{ob16} model, where the disc must have $Q\!<\!1$ in some parts to get $f_{\rm atm}\!\leq\!1$.  But the underlying physical motivation of both models is arguably the same, i.e.~unstable gas forms stars.  Indeed, the instability channel is the most dominant of the three star formation channels in \DS~\citepalias{sb17}.

\begin{figure*}
\centering
\includegraphics[width=\textwidth]{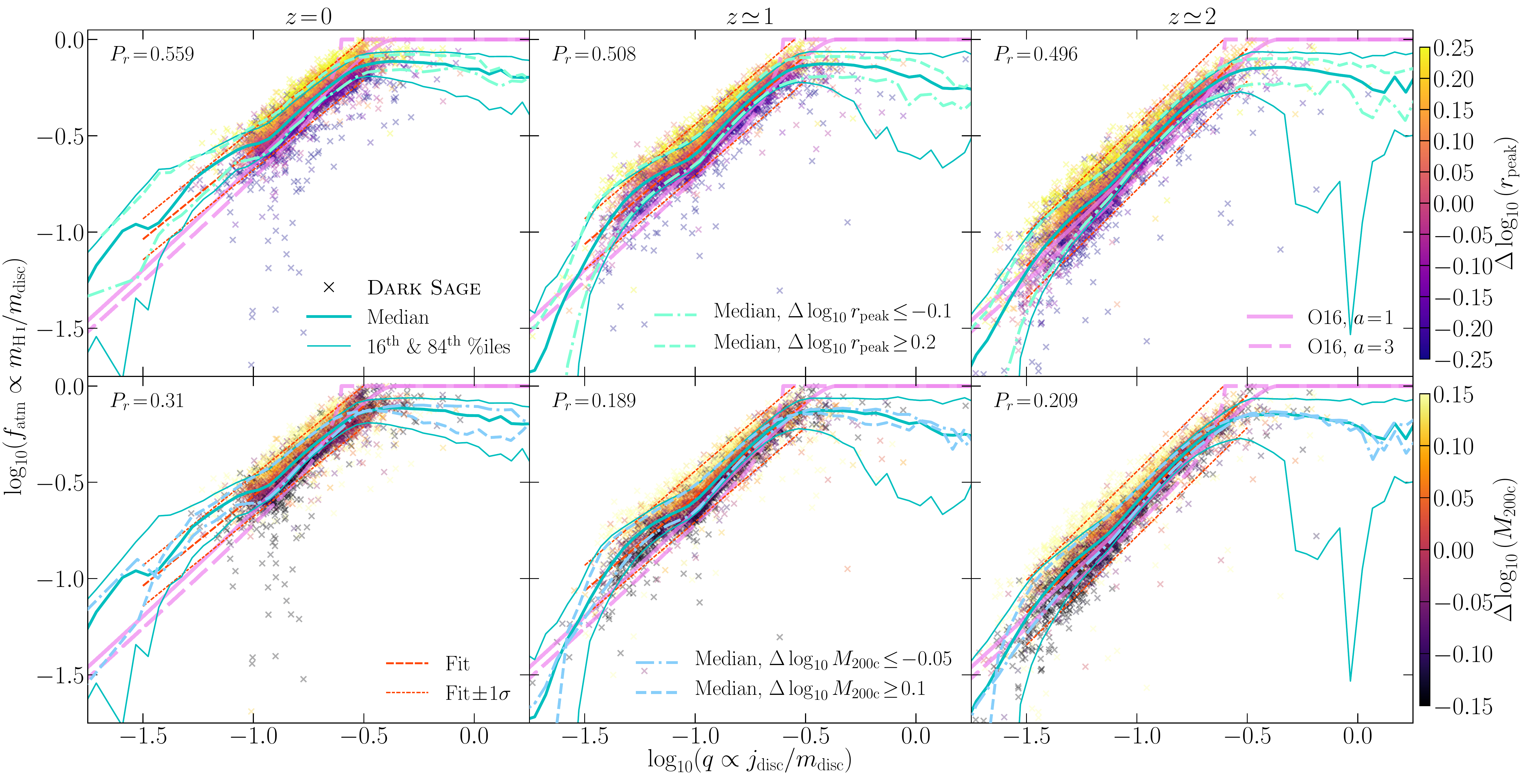}
\caption{Atomic gas fraction as a function of the `global disc stability' parameter (Equation \ref{eq:qfatm}b) for \DS~central, star-forming, disc-dominated galaxies' discs at three redshifts (one per column of panels).  Compared is the analytic model of \citet{ob16}.  4000 random individual \DS~galaxies are shown for each redshift.  Crosses in the top panels are coloured by the difference in radius where the probability distribution of $j$ peaks for that galaxy, compared to the median value of galaxies at the same $q$.  The bottom panels instead colour by the difference in halo mass from the median at fixed $q$. The Pearson-$r$ correlation coefficients, $P_r$, for $\Delta \log_{10}(f_{\rm atm})$ and the coloured properties are provided in the top-left of each panel.  In addition to running medians and percentiles for \DS~(built with bins of minimum width 0.05\,dex with at least 20 galaxies per bin), a linear fit over the range $-1.5 \!\leq\! \log_{10}(q) \!\leq\! -0.5$ is also shown; the standard deviation for galaxies' vertical displacement about this fit is included.}
\label{fig:q_fatm}
\end{figure*}

Fig.~\ref{fig:q_fatm} shows the $q$--$f_{\rm atm}$ relation for disc-dominated, central \DS~galaxies that are also star-forming (${\rm sSFR} \! \geq \! 10^{-11}\,{\rm yr}^{-1}$, defined consistently with previous \DS~papers) at three redshifts: 0, 1, and 2.%
\footnote{We note that while the median sSFR of galaxies systematically rises with redshift, we still use the same sSFR cut to define `star-forming' at each redshift.  This is because \DS~has multiple channels of star formation (see section 3 of \citetalias{stevens16} and figure 1 of \citetalias{sb17}).  The sSFR of galaxies forming stars predominantly through the instability channel changes by $\sim$1\,dex from \zo~to 2 (at fixed stellar mass), while only a factor of $\sim$2 change occurs for those forming through the passive H$_2$ channel.  We consider the latter to still be `star-forming', and thus find a cut of $10^{-11}\,{\rm yr}^{-1}$ appropriate at all $z \! \lesssim \! 2$.}  
Because the same cut is made at each redshift, the majority of galaxies in the $z\!\simeq\!2$ sample are \emph{not} progenitors of those at lower redshift.  The analytic model of \citetalias{ob16} overlies these \DS~galaxies remarkably closely at each redshift.  Consideration of satellites and quenched galaxies is left for Section \ref{sec:poor}.  One of the more notable differences in Fig.~\ref{fig:q_fatm} is that the sequence from \DS~galaxies has a shallower slope at $z\!=\!0$ and 1 than \citetalias{ob16}; the respective best least-squares linear fits in log-log space between $10^{-1.5} \!\leq\! q \!\leq\! 10^{-0.5}$ have gradients of 0.93 and 0.98, respectively.  At $z\!\simeq\!2$ though, the slope of 1.13 very closely matches the index of 1.12 from \citetalias{ob16}, as is the normalisation practically identical.  Another notable feature is the redshift evolution in the effective minimum $q$ for \DS~discs.  We come back to this point later in this section.  The flattening of the relation at high $q$ is simply a case of $f_{\rm atm}$ being unable to be greater than 1 by definition; discs with $q \! \gtrsim \! 10^{-0.5}$ are entirely stable, so a `global stability parameter' no longer helps to ascertain their atomic fractions (i.e.~the \citetalias{ob16} model ceases to be predictive here).

The scatter in the $q$--$f_{\rm atm}$ relation is where \DS~offers real physical insight over an analytic model.  We found many properties of the \DS~galaxies correlated with their vertical displacement from the median trend in the $q$--$f_{\rm atm}$ relation.  One of the strongest correlations (based on the Pearson-$r$ correlation coefficient) was with `total' disc mass, $m_{\rm disc}$.  One might expect this correlation to exist by construction, as $m_{\rm disc}$ is in the denominator of $f_{\rm atm}$ (see Equation \ref{eq:qfatm}).  The na\"{i}ve conclusion from this formula would be that scatter should be negatively correlated with disc mass.  Yet, instead, they are \emph{positively} correlated.  This is because, by definition, at fixed $q$, galaxies of high $m_{\rm disc}$ must also be of high $j_{\rm disc}$.  On average, higher $j$ means the disc will be less concentrated and therefore less efficient at converting atomic gas to molecular gas and, subsequently, stars.

\begin{figure*}
\centering
\includegraphics[width=\textwidth]{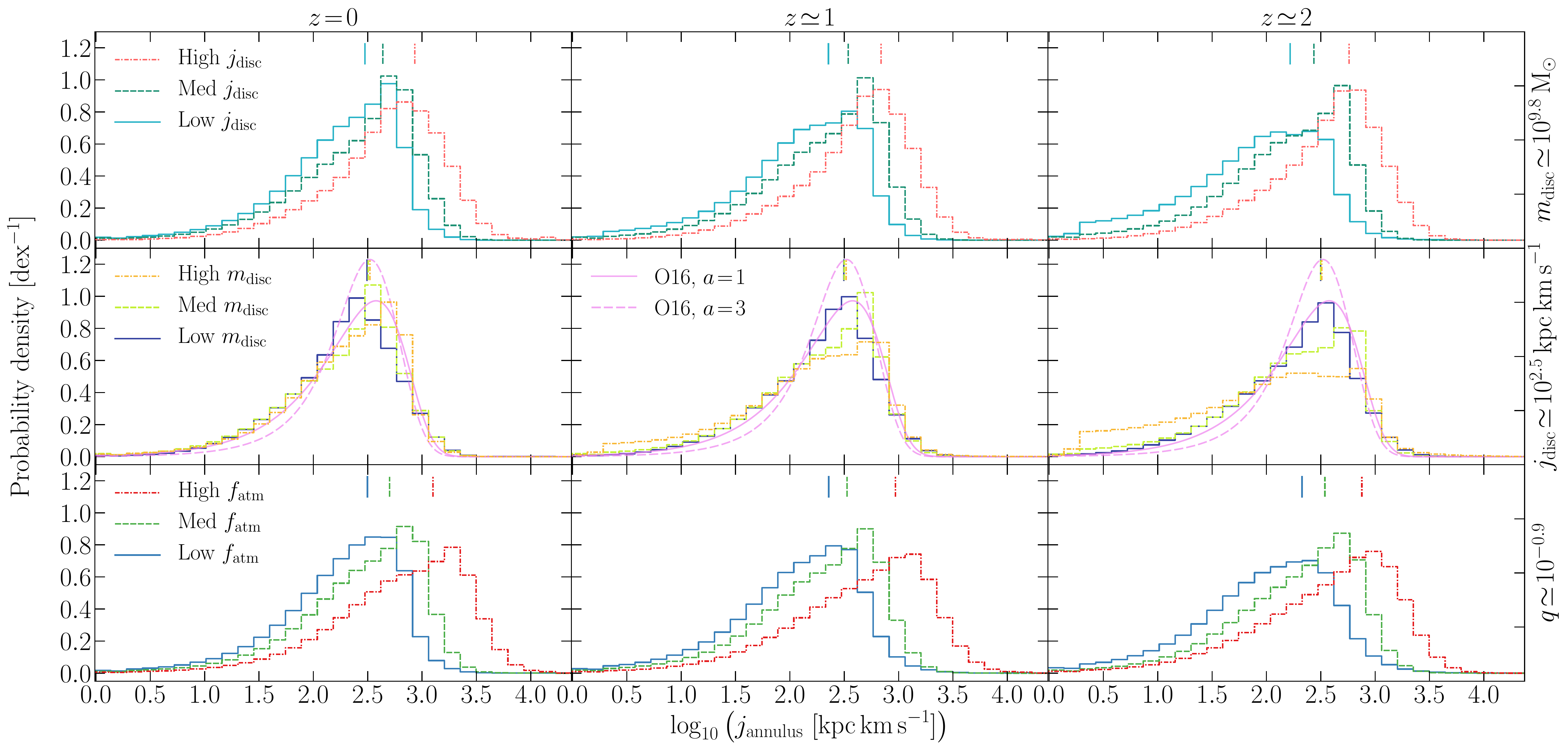}
\caption{Probability distribution functions of specific angular momentum within discs in \DS~galaxies.  Bins of $j$ are predetermined; they define the bounds between annuli in every galaxy, as per {\sc Dark Sage}'s design.  The top and middle rows of panels include galaxies within 0.05 dex of the quoted fixed disc mass and specific angular momentum on the right-hand side, respectively.  The bottom row includes galaxies within 0.01 dex of the quoted `global disc stability' parameter.  Each `low' histogram shows the mean \emph{normalised} value for each annulus for galaxies below the 10$^{\rm th}$ percentile of the quoted quantity.  Similarly, `med' PDFs consider galaxies between the 45$^{\rm th}$ and 55$^{\rm th}$ percentiles, and `high' ones are above the 90$^{\rm th}$ percentile.  Vertical dashes show the mean of each PDF of corresponding line style (equivalent to the mean $j_{\rm disc}$ of the galaxies contributing to that histogram).  The middle panels compare analytic PDFs of $j$ that match the assumptions of the \citet{ob16} disc model (also for $j_{\rm disc} \! = \! 10^{2.5}\,{\rm kpc}\,{\rm km}\,{\rm s}^{-1}$).}
\label{fig:j_pdf}
\end{figure*}

What the \citetalias{ob16} analytic model does not detail here is the variation between galaxies in the \emph{distribution} of $j$ \emph{within} a given disc.  This is limited by the assumption of  \citetalias{ob16} that discs have exponential surface density profiles and simple rotation curves; this translates into a common shape of the probability distribution function of mass a function of $j$ (PDF of $j$).  \DS, on the other hand, evolves both disc structure and rotation curves numerically, thereby predicting the PDF of $j$ within each disc.  Examples of these are presented in Fig.~\ref{fig:j_pdf} for the same redshifts as in Fig.~\ref{fig:q_fatm}.  These are simply calculated by summing (and normalising) the mass in common annuli for the relevant galaxy samples.  Compared are analytic PDFs of $j$ derived from the model assumptions of \citetalias{ob16} (which are not assumed to be redshift-dependent).

At fixed mass, \DS~discs with more total angular momentum naturally have PDFs of $j$ that are skewed towards (and therefore peak at) higher $j$ (top row of panels in Fig.~\ref{fig:j_pdf}).  At fixed $j_{\rm disc}$, the peak of the PDF also changes with $m_{\rm disc}$; to support the extra mass, a greater fraction of it must be at higher $j$.  That is, despite the means of the PDFs being the same, higher-mass discs peak at higher $j$ (middle row of panels) -- granted, this is (and can only be) a minor effect.  Now, when we consider galaxies at fixed $q$ and assess how the PDFs of $j$ vary with $f_{\rm atm}$, we have a compounding effect.  As noted above, the galaxies with higher $f_{\rm atm}$ have both higher $m_{\rm disc}$ and $j_{\rm disc}$, resulting in a clear trend between the peak of the PDF of $j$ and $f_{\rm atm}$ at fixed $q$ (bottom row of panels in Fig.~\ref{fig:j_pdf} -- the contribution from fixed $m_{\rm disc}$ notably outweighs that from fixed $j_{\rm disc}$).  This applies for all $q$, as can be seen from the colouring of each galaxy (cross) by the radius where its PDF of $j$ peaks, $r_{\rm peak}$, in the top panels of Fig.~\ref{fig:q_fatm} (in fact, it is the difference in the logarithm of this radius from the median of that of galaxies at the same $q$ that sets the colour).  Note that we have not coloured by the value of $j_{\rm peak}$, as this is discretised by construction in the model (the annuli in every galaxy are fixed in their specific angular momentum).  Because the conversion of $j$ to $r$ depends on the rotation curve of a galaxy, which varies on a galaxy-by-galaxy basis as per Equation (\ref{eq:j2r}), $r_{\rm peak}$ continuously varies between galaxies.  The fact that the 16$^{\rm th}$ percentile for all galaxies aligns so well with the median for galaxies with $\Delta \log_{10} r_{\rm peak} \! \leq \! -0.1$ at each redshift (and likewise the 84$^{\rm th}$ percentile for all with the median for $\Delta \log_{10} r_{\rm peak} \! > \! 0.2$) highlights that the peak of the PDF of $j$ truly is the primary cause of vertical scatter in the $q$--$f_{\rm atm}$ relation in \DS.

Fig.~\ref{fig:j_pdf} also highlights a clear redshift evolution in the PDFs of $j$ of \DS~discs. Both at fixed mass and fixed $q$, galaxies at lower redshift have systematically higher specific angular momenta (cf.~the vertical dashes that show the positions of the distribution means).  For fixed $q$, this presents itself as a shift of the entire PDF towards higher $j$ at low $z$.  For fixed $m_{\rm disc}$, it is less a case of the distribution peaks moving, and more a case of the PDFs becoming more peaked at low $z$.  These come from the fact that the lowest-$j$ material collapses into haloes and onto galaxies first, with higher-$j$ material following at later times \citep[\`{a} la][]{catelan96}.  For galaxies in \DS~this is implicitly coded in the connection of $j_{\rm cooling}$ to the halo (through Equation \ref{eq:rdnew}).  The net effect is a systematic increase in $j_{\rm disc} / m_{\rm disc}$ for all $m_{\rm disc}$.  The demand in \DS~that discs must be stable (that is, instabilities are resolved instantaneously, typically resulting in mass transfer from the disc to the bulge -- see section 3.7 of \citetalias{stevens16}) effectively places a cap on allowed disc masses (currently less than one per cent of \DS~discs exceed $10^{11}\,{\rm M}_{\odot}$, although the value of this pseudo-limit is set by the $w_{90}$ parameter -- see Appendix \ref{ssec:rs}).  Because $j_{\rm disc}$ at the maximum allowable disc mass increases with time, the effective minimum $q$ increases with time, which is seen in Fig.~\ref{fig:q_fatm} where the density of plotted crosses suddenly decreases.   As noted in Appendix \ref{ssec:rs}, the new $w_{90}$ parameter directly influences the value of this minimum $q$ (but does not significantly alter the other main features of Fig.~\ref{fig:q_fatm}), which has been calibrated to the observed disc-dominated galaxy stellar mass function (Appendix \ref{app:constraints}).


Because semi-analytic models are ultimately prescribed to evolve galaxies based on halo properties, we find it informative to relate the scatter in Fig.~\ref{fig:q_fatm} back to the halo.  Of the independent halo properties available in the merger trees, the most well correlated (based on its Pearson-$r$ correlation coefficient) with said scatter is $M_{\rm 200c}$ (which we refer to as the `virial mass' -- virial radius and velocity are obviously equally well correlated by definition).  With the update to cooling in \DS~(Appendix \ref{ssec:cooling}), the impact of halo specific angular momentum (and halo spin) on galaxy specific angular momentum has reduced.  However, the cooling scale radius still remains proportional to the virial radius in both the hot and cold modes of accretion.  Hence central galaxies in haloes of larger $M_{\rm 200c}$ will tend to have larger $j_{\rm disc}$, and this impacts scatter in $f_{\rm atm}$ at fixed $q$ more than $j_{\rm halo}$ or $\lambda$.  However, $j_{\rm disc}$ is determined by the halo's entire \emph{history}, not just its instantaneous value.  For $\log_{10}(q) \! \gtrsim \! -0.6$, $M_{\rm 200c}$ ceases to be connected to the scatter in $f_{\rm atm}$.  The absolute mass of haloes tends to decrease towards high $q$.  Haloes of lower mass typically have shorter and relatively variable histories, leading to incoherent growth, which detaches the hosted galaxy from the instantaneous halo properties.  At higher redshift, haloes have obviously had shorter histories, and consequently less-coherent growth.  As such, halo mass is a poorer indicator of scatter in $f_{\rm atm}$ at fixed $q$ (compare the closeness of the 16$^{\rm th}$ and 84$^{\rm th}$ percentiles with the medians for halo mass cuts in the bottom panels of Fig.~\ref{fig:q_fatm}). 

One thing we have not considered thus far is variations in the velocity dispersion of gas discs; \DS~assumes a constant $\sigma_{\rm gas} \! = \! 11\,{\rm km}\,{\rm s}^{-1}$.   In reality (observations), $\sigma_{\rm gas}$ varies both internally in a galaxy and between galaxies \citep[e.g.][]{tamburro09}.  For the \citetalias{ob16} model, this is inconsequential -- changing $\sigma_{\rm gas}$ for a galaxy would change both its $q$ and $f_{\rm atm}$ such that their relationship is preserved.  Instead, for \DS, we expect that allowing for variations in $\sigma_{\rm gas}$ would add compounding scatter to the $q$--$f_{\rm atm}$ relation.  This should not be a large effect though; any variation in \HI~velocity dispersion is observed to be much smaller than that for stars or H\,{\sc ii}, where the former is really the quantity of interest for $q$.

We have checked that the results presented in Fig.~\ref{fig:q_fatm} are not impacted by the annular resolution of \DS.  Specifically, we reran the model on a representative subset of the simulation trees with the same parameters after doubling the number of annuli in each disc, covering the same range in $j$, and came to the same conclusions.


\section{Breaking the connection: producing \HI-poor galaxies}
\label{sec:poor}
Here we investigate how galaxies can end up gas-poor for their $q$.  Indeed, galaxies in \DS~do populate the $q$--$f_{\rm atm}$ plane below the \citetalias{ob16} sequence.  These are (deliberately) not seen in Fig.~\ref{fig:q_fatm} though, as the galaxies in the \HI-poor area of parameter space are either satellites and/or quenched.

\subsection{Merger-induced quasar winds}
\label{ssec:mergers}
In examining galaxies that populate the entire $q$--$f_{\rm atm}$ parameter space, we find there is a distinct process in \DS~that can pull central galaxies off the sequence that applied for star-forming, disc-dominated galaxies in the model (Fig.~\ref{fig:q_fatm}).  That process is minor mergers.  Specifically, when a minor merger occurs, if both merging galaxies have non-zero gas disc masses, then colliding gas annuli are identified as those in the primary (larger baryonic mass) progenitor whose $j_{\rm annulus}$ is consistent with the orbital $j$ of the secondary progenitor about the primary just before it merged, convolved with a top-hat \citepalias[see section 3.9 of][]{stevens16}.  Within these colliding annuli, both direct accretion onto the central black hole and a starburst will be triggered \citep*[following modified versions of the respective prescriptions from][]{kauffmann00,somerville01}.  The former initiates quasar-mode feedback, which can unbind cold gas from disc annuli, potentially ejecting all the gas from the disc and the halo, depending on the quasar's energy (see \citealt{croton16}; \citetalias{stevens16}).  This can leave the galaxy devoid of \HI~and quenched, while the stellar component of the disc remains intact.  

The primary reason why major mergers do not have the same effect as minor mergers on the $q$--$f_{\rm atm}$ relation is that stellar discs  in \DS~are entirely destroyed during a major merger, with all stars transferred to the bulge.  While it is still possible for quasar activity (and/or the merger-driven starburst) to wipe out the gas disc in a major merger, the lack of stellar disc means there would no longer be a meaningful $q$ value to measure in such a case.  The disc would be entirely reset, growing like a new galaxy would, evolving once again along the main $q$--$f_{\rm atm}$ sequence.  In reality, mergers cannot be so harshly separated into two categories as they are in \DS~(and most semi-analytic models); the transition from minor to major merger should be continuous.  Nevertheless, it stands to reason that should a merger disrupt a galaxy's gas disc but not its stellar disc, $q$ and $f_{\rm atm}$ can become disconnected; whereas should both or neither be disrupted, $q$ and $f_{\rm atm}$ will remain causally tied.

\begin{figure}
\centering
\includegraphics[width=\textwidth]{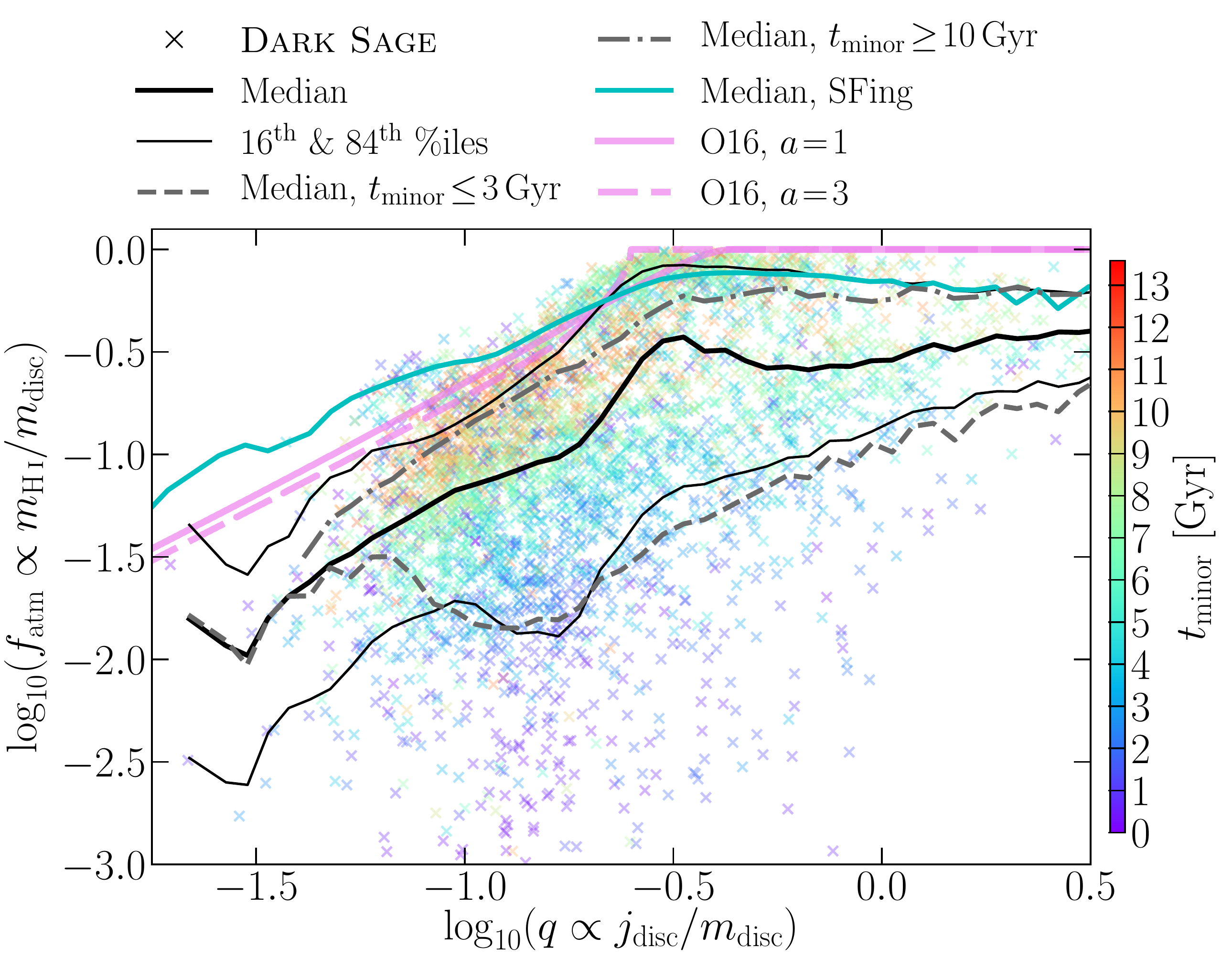}
\caption{`Global disc stability' parameter versus atomic gas fraction for quenched, disc-dominated, central \DS~galaxies at \zo~that have had at least one starburst-inducing minor merger, but have not had any major mergers.  Crosses are coloured by time since last minor merger.  Compared is the median trend for star-forming galaxies, i.e.~that copied directly from the left panels Fig.~\ref{fig:q_fatm}.}
\label{fig:qf_merge}
\end{figure}

In Fig.~\ref{fig:qf_merge}, we highlight galaxies that have gone through the above process with a minor merger.  We identify relevant galaxies at \zo~that are disc-dominated, with ${\rm sSFR} \! < \! 10^{-11}\, {\rm yr}^{-1}$, that have had at least one merger starburst, but have not had any major mergers.  The last criterion ensures that a \emph{minor} merger triggered said starburst(s), which will have been accompanied by quasar-mode feedback.  By colouring crosses for individual galaxies by their time since last minor merger, $t_{\rm minor}$, it is clear that those that have had a more recent merger tend to lie further from the $q$--$f_{\rm atm}$ sequence for star-forming discs.  Because the loss of \HI~from quasar winds is instantaneous in the model, galaxies are immediately maximally perturbed from this sequence when that happens.  From then on, the gas disc can slowly re-build through accretion/cooling.  In doing so, the galaxy evolves back towards the star-forming sequence, returning from `super-stability' to an equilibrium state of marginal stability.  The greater the value of $t_{\rm minor}$, the more opportunity a galaxy has had to return to this state, hence the trend seen in Fig.~\ref{fig:qf_merge}.  Galaxies will evolve back to the $q$--$f_{\rm atm}$ sequence at different rates (dependent on their cooling/accretion rates), so there is notable scatter in $f_{\rm atm}$ for galaxies of fixed $t_{\rm minor}$.  Additionally, in some cases, the quasar wind may only blow out the central disc gas, which is predominantly molecular, thereby leading to a minimal change in $f_{\rm atm}$, but still quenching the galaxy.

\begin{figure}
\centering
\includegraphics[width=\textwidth]{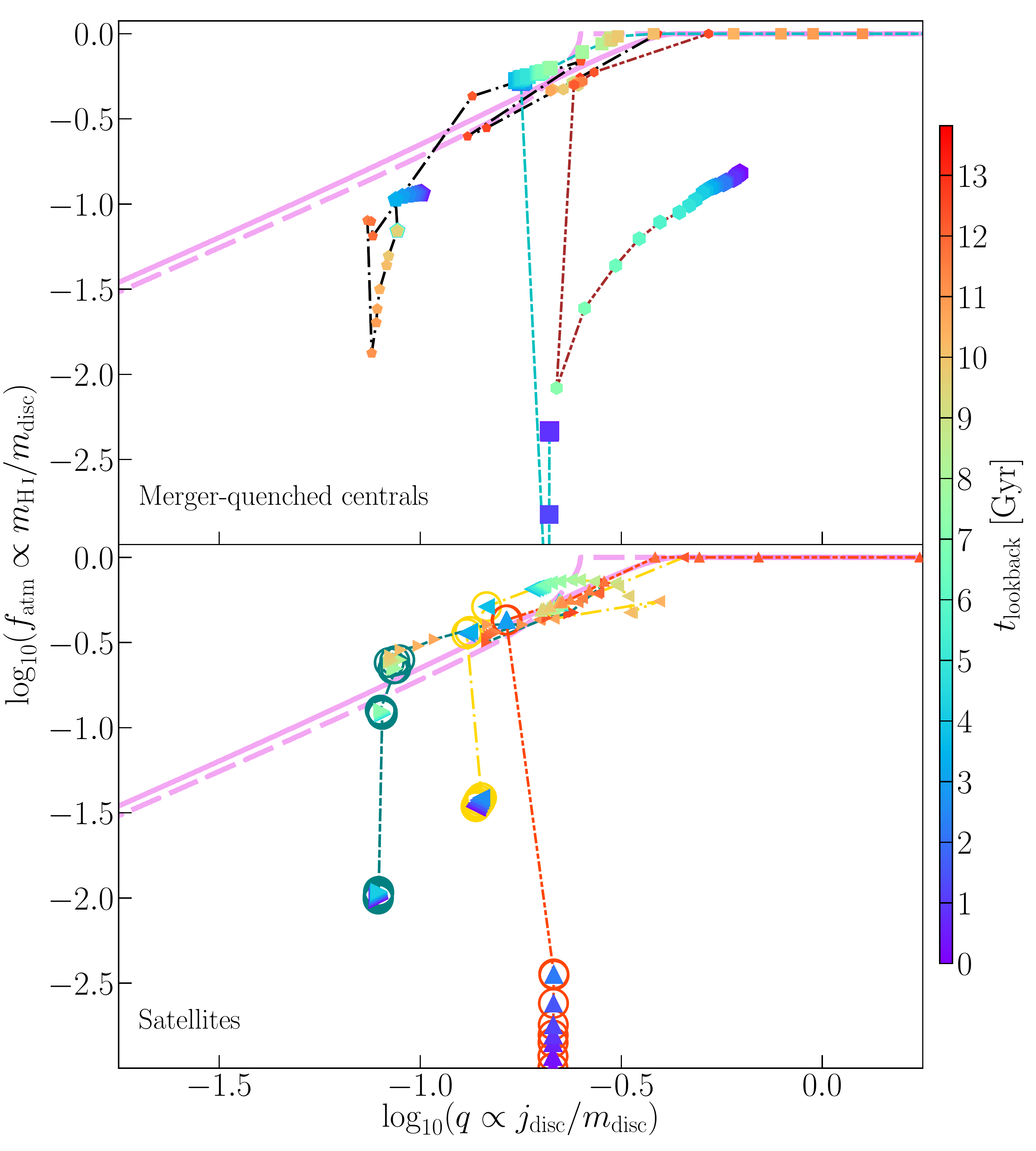}
\caption{Evolutionary tracks for example galaxies in terms of their `global disc stability' parameter and atomic gas fraction.  The top panel shows 3 instances of galaxies that suffer a minor merger that triggers a quasar, blowing out most of the galaxies' gas and quenching them.  The bottom panel shows 3 instances of galaxies that become satellites and have most of their \HI~stripped by ram pressure.  Different symbols and line-styles correspond to different galaxies.  Both the size and colour of the symbols indicate look-back time (larger means lower redshift).  Those that are circled in the bottom panel indicate that the galaxy was a satellite at that snapshot.}
\label{fig:qf_tracks}
\end{figure}

To highlight what happens to specific galaxies in more detail, we show evolutionary tracks for galaxies in $q$--$f_{\rm atm}$ space that have had quenching quasar activity triggered by a minor merger in the top panel of Fig.~\ref{fig:qf_tracks}.  Three semi-random examples are shown for galaxies that have different $q$ and $f_{\rm atm}$ values at \zo.  Normally, star-forming galaxies evolve from top right to bottom left along the $q$--$f_{\rm atm}$ sequence, as the rate of mass growth outpaces their rate of $j$ growth, and their \HI~fractions drop in the process.  After losing their gas from merger-induced quasar winds, after an almost vertical drop, they evolve in the opposite direction in parameter space (bottom left to top right), as the new gas they accrete is entirely high-$j$ (relative to the stellar disc) \HI~\citep[see][]{catelan96}.

It is worth noting that the process we have described here only applies to a small minority of \DS~central galaxies at \zo~that have had mergers: $\sim$6\% are quenched and have had a merger starburst(s).%
\footnote{This fraction accounts for bulge-dominated galaxies and those that have had both major and minor mergers.  For exclusively disc-dominated galaxies, $\sim$21\% are quenched and have had a merger starburst(s).  Further excluding those that have had major mergers, $\sim$32\% meet the criteria.}
The remainder lie relatively unperturbed on the star-forming $q$--$f_{\rm atm}$ sequence.  Controlled hydrodynamic simulations have found minor mergers can trigger substantial gas flow onto the central black hole(s), but this depends on both the orbital inclination of the secondary galaxy and its mass ratio to the primary \citep{younger08}.  In fact, earlier simulations without black holes found similar results; remnants from minor mergers ended up being gas-poorer for prograde orbits and higher mass ratios \citep*{bournaud05}.  It is, therefore, at least consistent that quasars are only triggered in a minority of \DS~cases.

For significant quasar activity to result from a minor merger in \DS~requires the colliding annuli of gas in the merger to constitute a significant fraction of the total combined cold gas of the progenitors.  This would maximise the black-hole growth and therefore the energy from quasar feedback to unbind the remaining gas in the disc.  For this to happen, the smaller progenitor must have a low orbital angular momentum, as this would lead to the inner annuli of the primary progenitor being classed as the collisional ones; because annuli are spaced logarithmically, this would maximise the \emph{number} of collisional annuli, and hence maximise the mass of `collisional gas.'  While there is a parallel to be drawn with the above-described results of hydrodynamic simulations that the orbital parameters of the secondary progenitor are key to the black-hole growth, \DS~is less concerned with the secondary having had a pro- or retrograde orbit, and more with the magnitude of its angular momentum component projected onto the primary's disc plane.

At this point, it is worth stepping back to reassess how well physically motivated the treatment of black-hole growth and quasar-mode feedback is in \DS~(specifically for the case of minor mergers).  This is one aspect of the model that underwent minimal development from the original version of {\sc sage} \citep{croton16}, which followed \citet{kauffmann00}.  In this framework, the fraction of cold gas accreted by the black hole scales linearly with the secondary-to-primary baryonic mass ratio of the progenitor galaxies.  \DS~simply modifies this to apply for each colliding annulus individually, where the fraction of gas in that annulus that is accreted onto the black hole scales with the ratio (always $<$1) of gas in that annulus originating from the primary and secondary progenitors.  This modification was made to make the prescription relevant for the design of \DS~discs, but a physical argument for it was not proposed by \citetalias{stevens16}.  One potentially important difference introduced is that the stellar content of merging galaxies no longer impacts the black-hole accretion rates.  Based on the large scatter in galaxy gas masses at fixed stellar mass \citepalias{sb17}, reintroducing this dependence should, in principle, reduce the black-hole feeding rates from collisional gas annuli with comparable mass contributions from both progenitors.

One may also argue that, physically, the gas that feeds the black hole should first flow through the disc, rather than feed it directly as in \DS.  Of course, the entire process of mergers is not instantaneous, but the nature of how semi-analytic models are designed, i.e.~that galaxies are always in some axisymmetric equilibrium state at each snapshot, unfortunately means mergers are resolved instantaneously.  Arguably, the gas that feeds the black hole in \DS~\emph{does} flow through the disc (quickly), but simply does not entrain lower-$j$ gas in doing so.  As \citet{capelo15} note, it is not necessarily the lowest-$j$ gas in a merger that feeds the black hole, but rather the gas that \emph{loses} $j$.

Certainly, there is a significant body of observational evidence that outflows of cold gas are tied to active galactic nuclei \citep[e.g. see reviews by][]{heckman14,king15}. But based on observations, it is not entirely clear how often quasar winds should be capable of ejecting the majority of a galaxy's gas reservoir, or how quickly they might manage to do so.  
While the result of Fig.~\ref{fig:qf_merge} is a genuine prediction of \DS, because of the imprecise manner in which quasar-mode feedback is prescribed in the model, this prediction should be taken more qualitatively than quantitatively.  If a similar population of quenched, low-$f_{\rm atm}$, disc-dominated central galaxies could be identified in other models or cosmological hydrodynamic simulations (which are distinct from semi-analytics in their methodology for how the properties of galaxies are built up), this prediction would have firmer foundations.  We discuss what evidence currently exists in observations in Section \ref{ssec:observing}, including debate surrounding the connection between mergers and quasar activity.

\subsection{Ram-pressure stripping of satellites}
\label{ssec:rps}
\begin{figure}
\centering
\includegraphics[width=\textwidth]{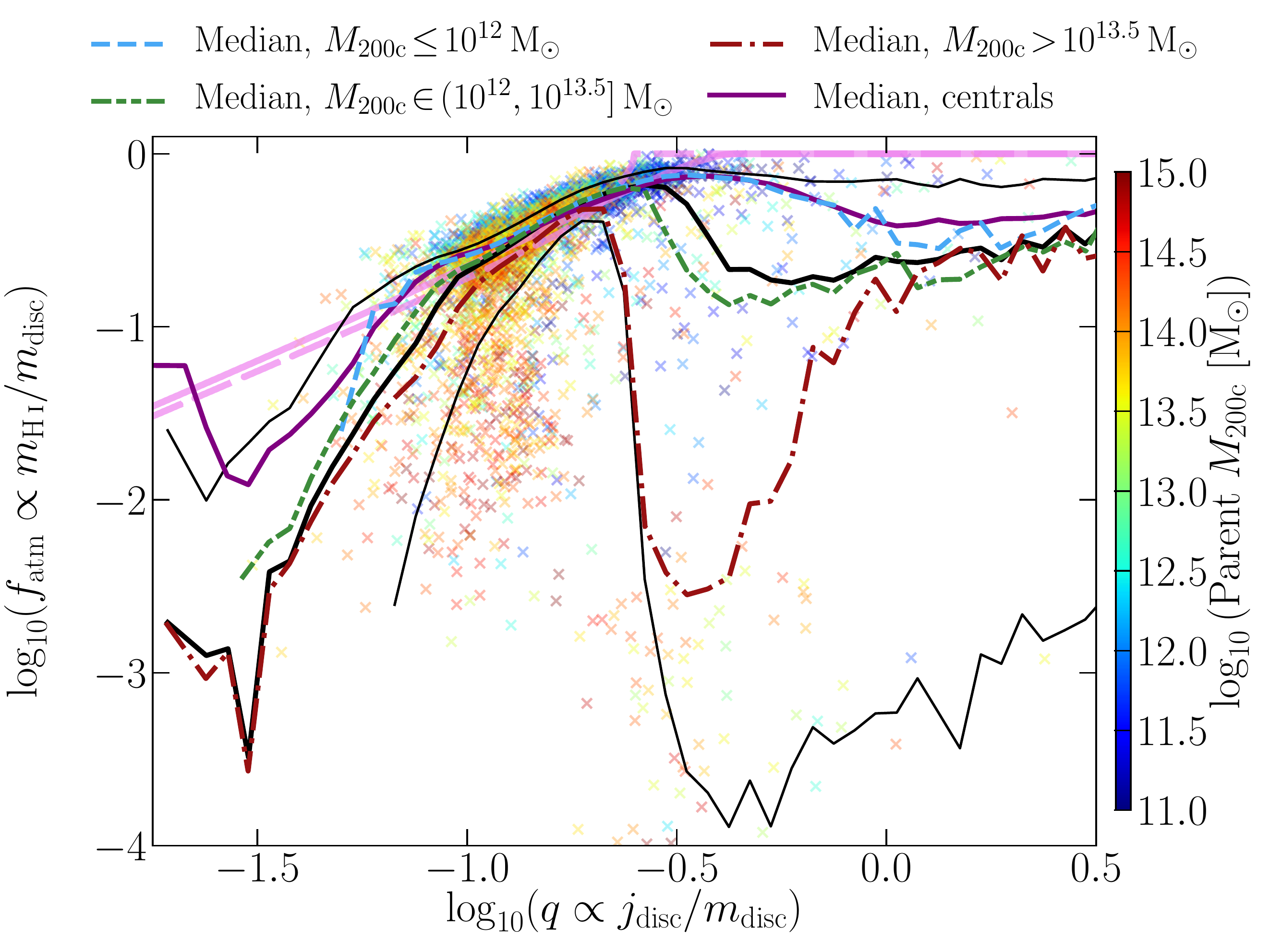}
\caption{As for Fig.~\ref{fig:qf_merge}, except now showing disc-dominated, satellite \DS~galaxies, with points coloured by the main halo mass that the satellites are embedded in.  Galaxies are broken into three bins of halo mass, with the median for each bin shown, as per the legend.  For each of these bins, 1200 random individual galaxies are shown.  Compared is the median for \emph{all} disc-dominated centrals (i.e.~both quenched and star-forming galaxies, which differs from Fig.~\ref{fig:q_fatm}).}
\label{fig:qf_sat}
\end{figure}

In Fig.~\ref{fig:qf_sat}, we present the $q$--$f_{\rm atm}$ relation for all disc-dominated satellite galaxies in \DS~at $z\!=\!0$.  By directly comparing the median from these galaxies against equivalent centrals, we find that satellites and centrals occupy the same sequence for $10^{-1} \! \lesssim q \! \lesssim 10^{-0.6}$.  Outside this range, satellites are relatively gas-poor.  By further comparing medians for satellites in three bins of parent halo mass, it is clear that environment is affecting the satellites; those in more massive haloes have lower \HI~fractions for their $q$.

The aspect of \DS~that pulls satellites off the $q$--$f_{\rm atm}$ relation is ram-pressure stripping of the cold disc gas.  Gas stripping in \DS~is a two-stage process.  Initially, only the satellite's hot gas is susceptible to stripping.  In the default implementation (applied in this work), hot-gas stripping follows the rate of dark-matter loss in the merger trees (which can generally be attributed to tidal stripping) multiplied by the cosmic baryon fraction (for a comparison to a ram-pressured-based prescription, see \citetalias{sb17}).  Once a sufficient amount of hot-gas mass is lost (for \DS, this is when the hot-gas mass becomes less than the combined mass of cold gas and stars in the satellite galaxy), the cold gas is no longer considered to be protected by the hot gas, and thus cold-gas stripping is activated.  For each annulus in the cold-gas disc, the ram pressure felt by the satellite (based on its relative velocity to the central and the expected density of the intra-halo gas at its position) is compared to the local gravitational restoring force \citep{gunn72}.  Should ram pressure win, all gas in that annulus is transferred to the hot-gas reservoir of the central.  Otherwise, the annulus is left unperturbed.  For further details, see section 2.1 of \citetalias{sb17}.

The natural outcome of our ram-pressure implementation is that the outer, lower-density annuli are stripped first.  Gas in those annuli is predominantly atomic (and generally not star-forming).  Thus, as highlighted in detail by \citetalias{sb17}, ram-pressure stripping of the cold gas disc in \DS~leads to a distinct trend between \HI~fraction and parent halo mass at fixed stellar mass, which matches observations to a more accurate degree than previous model results in the literature \citep[also see][]{brown17}.  This finding is easily related to ram pressure being stronger in denser environments.  With Fig.~\ref{fig:qf_sat}, we are seeing the same effect, just at fixed $q$ instead.  It is interesting to note that while halo mass clearly has an impact on satellites' \HI~fractions for all $m_* \! > \! 10^9\,{\rm M}_{\odot}$, the signature is only strong over a narrow range in $q$, approximately $\left[10^{-0.6},10^{-0.1}\right]$.  Such high values of $q$ imply two things.  First is that these are predominantly low-mass galaxies; if one takes the nominal scaling of $j \! \approxprop \! m^{2/3}$ \citep[see][and references therein]{rf12}, then one finds $q \! \propto \! j/m \! \approxprop \! m^{-1/3}$.  These low-mass satellites are extremely susceptible to having their \HI~stripped by their environment, as they lack the gravitational potential to prevent ram-pressure stripping.  Secondly and alternatively, high $q$ can also imply moderate mass but \emph{relatively} high $j$, meaning those galaxies have lower average surface densities, which again makes them (or, rather, their outer regions) more prone to stripping.  

Satellites of $q \! \gtrsim \! 1$ are sufficiently low-mass such that they are less likely to have been a satellite for a long period of time; \DS~does not include orphan galaxies, so as soon as tidal stripping causes a subhalo to drop below 20 particles, the satellite is immediately either disrupted or merged \citep[see][]{croton16}.  Also, these galaxies are so susceptible to having their extended \HI~stripped, even low-mass haloes can do so efficiently.  As such, the trend with halo mass for $f_{\rm atm}$ at fixed $q$ is lost.  Note that the vast majority of satellites still have \emph{some} \HI~in their centres; only $\sim$5\% of satellites that contribute to Fig.~\ref{fig:qf_sat} have zero \HI.

At $q \! \lesssim \! 0.1$, halo mass can again be seen to impact \HI~fractions, but more weakly than at high $q$.  Again, taking $j \! \approxprop \! m^{2/3}$, one expects $q \! \approxprop \! j^{-1/2}$.  This means galaxies at low $q$ tend to have higher specific angular momenta, and hence are typically larger in absolute units.  Their outer-most gas is therefore susceptible to stripping, but significant gravitational potentials limit the impact of stripping.  At intermediate $q$, \DS~satellites are almost unaffected by their environment, as they typically have sufficiently deep potential wells and are sufficiently compact to protect themselves from ram-pressure stripping.

The bottom panel of Fig.~\ref{fig:qf_tracks} shows evolutionary tracks for three example cases of galaxies that become satellites and suffer ram-pressure stripping.  In each case, the galaxies drop in their \HI~fraction relatively rapidly soon after infall, with only a minor change to their $q$.  Low-$q$ (high mass) galaxies have a longer delay before this drop takes place, as they have larger protective hot-gas atmospheres that must first be stripped.  After this, in the two lower-$q$ cases, both galaxies only show minor decreases to their \HI~content.  The slow evolution indicates that these galaxies have either already passed their point of maximum ram pressure or their only remaining gas is at sufficiently high density (in low-$j$ annuli) such that it is not affected by additional ram pressure.  The higher $q$ case shows a clearer continuation of gradual \HI~stripping.

\subsection{On observing these effects}
\label{ssec:observing}
\begin{figure*}
\centering
\includegraphics[width=0.7\textwidth]{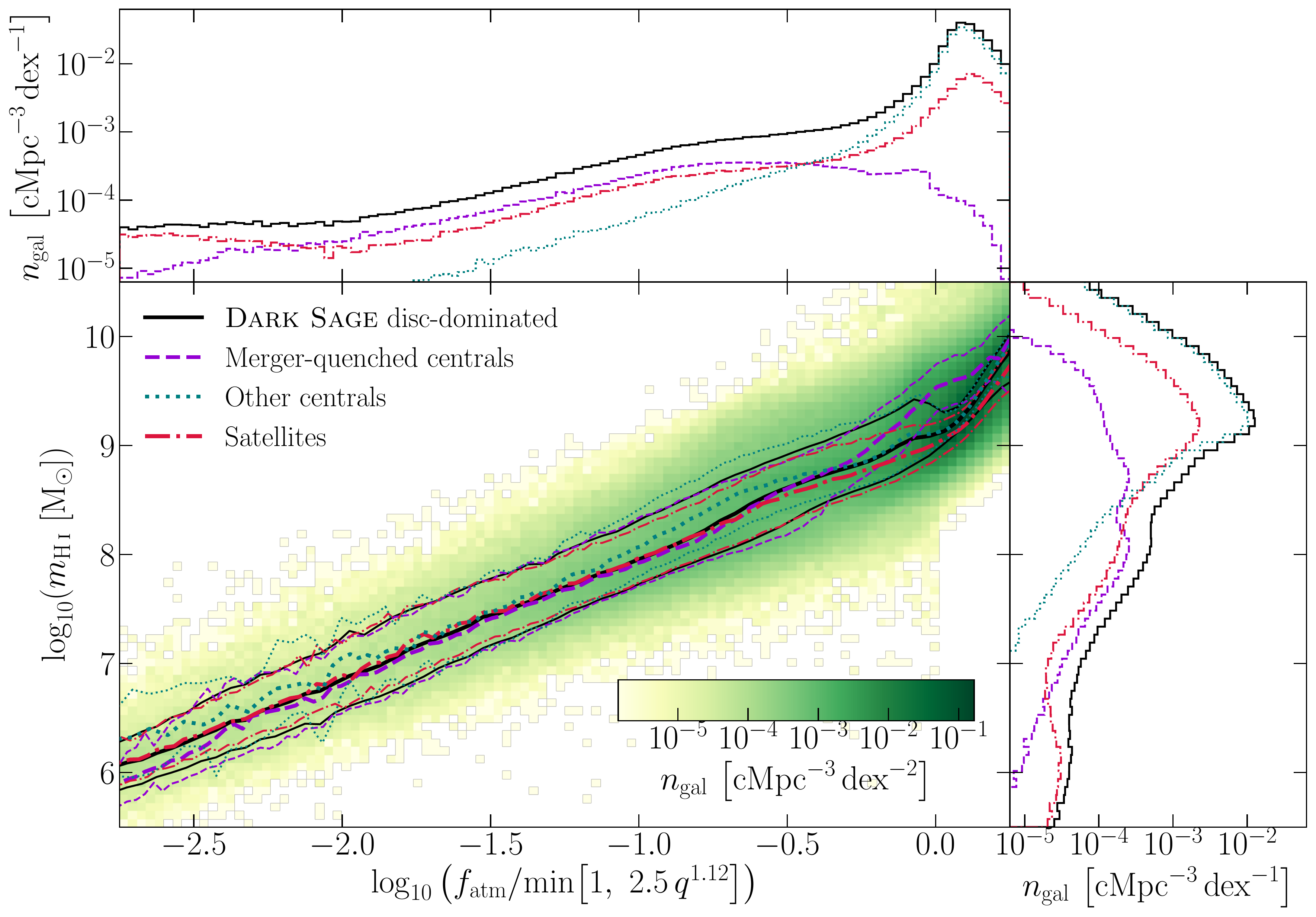}
\caption{Absolute \HI~mass of galaxies in terms of their \HI-fraction difference from the \citetalias{ob16} model at \zo.  Thick lines in the main panel are medians.  Thin lines are 16$^{\rm th}$ and 84$^{\rm th}$ percentiles.  Compared against the complete population of disc-dominated \DS~galaxies are the subsamples from Sections \ref{ssec:mergers} (centrals quenched from quasar activity triggered by a minor merger) and \ref{ssec:rps} (satellites, which are subject to ram-pressure stripping).  Summing these subsamples with the `other centrals' gives the full disc-dominated sample.}
\label{fig:deficit_HI}
\end{figure*}

To test the predictions from Sections \ref{ssec:mergers} and \ref{ssec:rps}, one needs to observe a sufficient sample of galaxies that occupy the region of $q$--$f_{\rm atm}$ parameter space that is below the \citetalias{ob16} relation (equivalent to the star-forming sequence from \DS).  A thorough study would require observing a statistically significant sample of galaxies extending to at least 1\,dex below the analytic relation, over an order of magnitude in $q$, and covering a variety of environments.  One of the main challenges here is having \HI~detections for those galaxies, as by definition, they must have low absolute gas masses.  To highlight this, we show in Fig.~\ref{fig:deficit_HI} what the typical absolute \HI~masses of \DS~galaxies are for deficits in $f_{\rm atm}$.  By comparing all \DS~disc-dominated galaxies against the subsamples assessed in Sections \ref{ssec:mergers} and \ref{ssec:rps}, we note that the cause of why the galaxies are \HI-poor according to their $q$ is irrelevant when it comes to their absolute mass, and hence the challenges associated with observing the relevant galaxies (at least from an instrumental perspective) are the same.  The average number densities of galaxies that are 0.5 or more dex deficient in \HI~for their $q$ are approximately equal for those stripped by ram pressure and those depleted by quasar feedback from minor mergers (as seen by the one-dimensional histograms in the top and right panels of Fig.~\ref{fig:deficit_HI}).

To comfortably detect galaxies with a 1-dex deficiency in $f_{\rm atm}$ would require a survey(s) with \HI~detections at $m_{\rm H\,{\LARGE{\textsc i}}} \! \leq 10^8\,{\rm M}_{\odot}$.  \HI-blind surveys like ALFALFA\footnote{Arecibo Legacy Fast Arecibo $L$-band Feed Array} have detected a small fraction of their tens of thousands of galaxies below $10^8\,{\rm M}_{\odot}$ in \HI~within $\sim$40\,Mpc (\citealt{haynes11}; also see \citealt{jones18}).  Unfortunately, the nature of blind surveys means these will be biased towards higher \HI~masses.  While ongoing surveys such as {\sc wallaby} will significantly boost statistics of like galaxies once completed \citep[for details, see][]{duffy12,koribalski12}, one ideally requires more-targeted observations that can probe lower \HI~masses.  The VIVA survey \citep{chung09}, for example, is a prime candidate for examining the impact of ram pressure at fixed $q$ (Section \ref{ssec:rps}), with \HI~detections down to nearly $10^7\,{\rm M}_{\odot}$, corresponding to $f_{\rm atm}$ deficits of $\sim$2 dex (based on Fig.~\ref{fig:deficit_HI}).  This will be assessed in an upcoming paper by J.~Li et al.~(in preparation).  One would need to examine more than one cluster to draw generic conclusion though.

With that said, there exist recent published results from \HI~surveys that are apropos to our predictions.  Using data from the xGASS survey \citep{catinella18} plus additional data that followed the same observing strategy, \citet*{ellison18} find that (central) merger remnants are systematically more gas-\emph{rich} than their control counterparts of the same stellar mass.  While this might sound contrary to our results in Section \ref{ssec:mergers}, there was no explicit consideration of the galaxies' specific angular momenta in their work, so whether they are \HI-poor for their $q$ remains open. 
\citet{ellison18} note that, based on simulations by \citet{lotz10}, their selection might favour mergers whose progenitors had high gas fractions, suggesting they could be remnants of wet mergers.  While no explicit comment is made on their being major or minor mergers, if they were wet minor mergers, then, based on the EAGLE results of \citet{lagos18}, we would expect the merger remnants to have systematically higher $j$ for their mass, suggesting their high gas fractions may simply be a result of higher $q$ \citep[see][]{lutz18}.  Given that many of these could be major merger remnants as well, it is likely that the majority of their sample is \emph{not} analogous to a subsample of the \DS~galaxies presented in Section \ref{ssec:mergers}.  Nevertheless, it is interesting to note that \citet{ellison18} find no connection with active galactic nuclei in the merger remnants and their gas content.
In a different, relevant study, based on ALFALFA data and subsequent Arecibo observations, \citet{bradford18} find that over a narrow mass range $m_*/{\rm M}_{\odot} \! \in \! \left[10^{9.2},\, 10^{9.5}\right]$, there is a population of \HI-depleted isolated galaxies in the local Universe that are quenched and show signs of AGN activity.  Galaxies like these could be candidates for the post-minor-merger scenario we have described in this paper.

Beyond \HI~masses, another significant challenge for testing our above predictions is measuring accurate values of $j_{\rm disc}$ for galaxies.  Only with resolved structural information of the galaxies' mass distributions and kinematics can this be done \citepalias[e.g.][]{og14}.  The telescope time required for these data for a large sample of galaxies is immense.  While {\sc wallaby} should go some way towards this, ultimately, assumptions will need to be made regarding galaxy structure, rotation, and dispersion to estimate $q$ (see \citetalias{ob16}; \citealt{lutz18}).  This adds a layer of uncertainty, which will likely introduce systematics and increase observed scatter between $q$ and $f_{\rm atm}$.


\section{Summary}
\label{sec:summary}

We have updated the \DS~semi-analytic model \citepalias{stevens16,sb17}, which numerically evolves the one-dimensional structure of galactic discs in annuli of fixed specific angular momentum, with three new features, described in Appendix \ref{app:updates}.
%
%
%
%
The model has been recalibrated to meet a series of observational constraints, as presented in Appendix \ref{app:constraints}.  The updated codebase is publicly available through the \DS~Github repository.$^{\ref{foot:github}}$ The model is also listed in the Astrophysics Source Code Library.\footnote{\url http://ascl.net/1706.004}

With \DS, we have investigated the connection between disc specific angular momentum, mass, and gas fraction.  We have highlighted the distinct correlation between gas fraction and $j$ for fixed disc mass when considering stellar content (Fig.~\ref{fig:jm_gasfrac}, which is in line with recent results from observations and other models: \citetalias{og14}; \citealt{lagos17}; \citealt{zoldan18}), as well as the stronger connection when considering all baryons in the disc (Fig.~\ref{fig:jm_gasfrac_bary}).  These results have been framed alongside the analytic model of \citetalias{ob16}, who theorise that this connection arises from specific angular momentum setting the `global disc stability' parameter, $q$, of a hypothetical, single-component, warm, neutral disc, which in turn sets the ratio of baryons in the disc that will remain in an atomic, gaseous state, $f_{\rm atm}$ (Section \ref{sec:causal}).

In using a semi-analytic model with predictive disc structure, we have extended the connection between $q$ and $f_{\rm atm}$ to now include hierarchical assembly and galaxy evolution physics. Central, star-forming, disc-dominated galaxies in \DS~produce a clear sequence between $q$ and $f_{\rm atm}$, which is remarkably in line with the model of \citetalias{ob16}.  Where \DS~excels is in highlighting the scatter in this relation, which we have shown is caused by variations in how $j$ is distributed within discs (Figs.~\ref{fig:q_fatm} \& \ref{fig:j_pdf}).  Variations in $f_{\rm atm}$ at fixed $q$ can be almost entirely explained by the radius where the probability distribution function of $j$  in the disc peaks.  To first order, this is driven by halo mass, but the correlation between this and variations in $f_{\rm atm}$ only applies at low $q$, where haloes are larger and have more stable histories.  These conclusions apply for all $z\!\lesssim\!2$.  We did not assess higher redshift, as assumptions in the model regarding disc formation break down there.

While the $q$--$f_{\rm atm}$ relation is tight for star-forming central galaxies, this is not the case for either quenched centrals or satellites galaxies in dense environments.  Minor mergers in \DS~can lead to rapid black-hole growth and subsequent quasar-mode feedback that can blow out the cold gas in the disc entirely, while leaving the stellar disc intact (Section \ref{ssec:mergers}).  This causes galaxies to rapidly drop in $f_{\rm atm}$, while maintaining a similar $q$.  As these galaxies start to accrete gas again, they slowly evolve back towards the star-forming sequence.  It remains somewhat unclear if this prediction is an artefact of the model's design.  While updates to \emph{radio}-mode growth and feedback of black holes from another {\sc sage} branch are already published \citep{raouf17}, future work is required to update the treatment of quasar-mode feedback in \DS.  Satellite galaxies can also become \HI~deficient for their $q$ as a result of ram-pressure stripping (Section \ref{ssec:rps}).  We note that testing these predictions observationally with statistical rigour will be challenging, both because of the absolutely low \HI~masses of these galaxies and the extended kinematics and multi-wavelength information necessary to accurately measure $j$ (and hence $q$).  In principle, surveys like xGASS, VIVA, and {\sc wallaby} should be able to help provide insight -- in the case of xGASS, this has already begun (Section \ref{ssec:observing}).

Hydrodynamic simulations offer another avenue of investigation into the $q$--$f_{\rm atm}$ connection and the processes that affect it.  Challenges naturally come with the deconstruction of gas elements into their ionized, atomic, and molecular components, as well as in separating disc, bulge, and intrahalo contributions of each galaxy, although progress in these areas has been made \citep[e.g.][]{stevens14,lagos15,canas18,diemer18,mitchell18}.  Works by L.~Wang et al.~(in preparation) and J.~Li et al.~(in preparation) will assess the $q$--$f_{\rm atm}$ relation in the NIHAO \citep{wang15} and EAGLE simulations \citep{schaye15}, respectively, with the latter focussing on the effect of environment on satellites.  Complementarily, using the IllustrisTNG simulations \citep{pillepich18}, we will also explore the impact of environment on satellites (A.~R.~H.~Stevens et al., in preparation).


\section*{Acknowledgements}
All plots in this paper were built with the {\sc matplotlib} package for {\sc python} \citep{hunter07}.  ARHS thanks S.~Bellstedt, B.~Catinella, D.~Fisher, K.~Glazebrook, J.~Lie, S.~Sweet, and L.~Wang for productive discussions related to this work.  CL is funded by an Australian Research Council Discovery Early Career Researcher Award (DE150100618) and by the Australian Research Council Centre of Excellence for All Sky Astrophysics in 3 Dimensions (ASTRO 3D), through project number CE170100013.

\appendix

\section{New model features}
\label{app:updates}

\subsection{Update to hot-mode cooling scale radii}
\label{ssec:cooling}
When gas cools onto a galaxy in {\sc Dark Sage}, it is distributed with a surface density that exponentially declines with specific angular momentum:
\begin{equation}
\Sigma_{\rm cool}(j) = \frac{\Delta m_{\rm cold}}{2 \pi r_{\rm d}^2} \exp\left(\frac{-j}{r_{\rm d} V_{\rm vir}}\right)\,.
\label{eq:cooling}
\end{equation}
The orientation of $\hat{j}_{\rm cooling}$ matches that of the halo at that time.  The angular-momentum vector of the cooling gas is summed with that of the pre-existing gas disc to define the disc's new plane, where any orthogonal component of angular momentum to that plane is dissipated \citepalias[see section 3.3 of][]{stevens16}.  

The departure from a nominal exponential decline with \emph{radius} \citep[\`{a} la][]{fall80} in Equation (\ref{eq:cooling}) fits with the model's explicit consideration of discs in angular-momentum space.  Note, though, that there is still an explicit dependence on a cooling scale radius, $r_{\rm d}$.  Originally, this followed the commonly adopted linear scaling with the halo spin parameter, $\lambda$ \citep[see][]{fall83,mo98}.  Recent analysis of the EAGLE simulations by \citet{stevens17} has shown that the best-fitting scale radius of gas that cools onto galaxies departs from a \emph{linear} scaling with $\lambda$, at least when galaxies are accreting in the hot mode.  In light of this, we have updated the cooling scale radius in \DS~to follow the fitting function of \citet{stevens17}:
\begin{equation}
\log_{10}\left(\frac{r_{\rm d}}{R_{\rm vir}}\right) = m_{\rm d} \log_{10}\lambda - k_{\rm d}~,
\label{eq:rdnew}
\end{equation}
where $m_{\rm d}$ and $k_{\rm d}$ were fitted parameters.  In this work, we fix $m_{\rm d} \! = \! 0.23$, as per the best-fitting value for the high-resolution EAGLE simulation in equation 19 of \citet{stevens17}.  We do not, however, directly adopt the $k_{\rm d}$ value from the same equation.  This is because (a) that value was not converged with the simulation resolution, and more importantly, (b) this included gas that cooled from outside the virial radius of the halo; \DS~assumes all hot gas, and therefore all cooling gas, is within the virial radius.  \citet{stevens17} note that if only cooling gas within the virial radius is considered, then $\langle j_{\rm cooling} / j_{\rm halo} \rangle \! \simeq \! 1.4$.  Because cooling gas is distributed according to Equation (\ref{eq:cooling}), this ratio translates to $\langle r_{\rm d} / r_{\rm d,old} \rangle  \! \simeq \! 1.4$.  To recover this, we set $k_{\rm d} \! = \! 1.0$.

Equation (\ref{eq:rdnew}) \emph{only} applies for hot-mode accretion; for the cold mode, we maintain the same $r_{\rm d}$ as in \citetalias{stevens16}.  This is because the results of \citet{stevens17} did not include galaxies accreting in the cold mode.  Several other studies of hydrodynamic simulations have found similar results regarding the angular momentum of gas in galaxies and haloes accreting in the cold mode though \citep[e.g.][]{kimm11,stewart11,danovich15}.  This gives grounds to extend Equation (\ref{eq:rdnew}) or an equivalent to cold-mode accretion in \DS.  We leave this as a point of investigation for future work.

We also note that three other findings of \citet{stevens17} were that (i) cooling gas is better represented as being distributed exponentially with radius rather than $j$, (ii) gas loses $\sim$60 per cent of its angular momentum during cooling, and (iii) different feedback implementations in EAGLE had little effect on where and when gas cooled.  Because disc content of a given $j$ is assumed to be radially co-located for an individual \DS~galaxy (which need not be true in real galaxies or those in hydrodynamic simulations), the first point is, in principle, not of significant consequence.  We leave this as an available option in the model though, and find it only demands minor recalibration for our chosen constraints (see Section \ref{ssec:recalibration}).  As for the second point, while we have not modified \DS~to explicitly account for this loss in $j$, the projection of ${\bm j}_{\rm cooling}$ onto the updated disc plane (as described above) does already account for some loss of $j$ during cooling.  The third point helps validate the use of Equation (\ref{eq:rdnew}) in \DS, suggesting it is not an artefact of the subgrid physics implemented in EAGLE.

\subsection{Update to stellar disc scale radii}
\label{ssec:rs}
In the earlier versions of {\sc Dark Sage}, no distinction was made between the scale radius for cooling and the actual scale radius of a galaxy's stellar disc for the purposes of semi-analytic prescriptions (that needed to interpret the disc `size').  We now introduce that distinction to \DS.  As defined below (Equation \ref{eq:rs}), a stellar disc scale radius, $r_{\rm s}$, is routinely calculated in the model, which updates the exponential scale radius for the stellar velocity dispersion profile and for informing the scale radius of the instability-driven bulge (i.e. equations 11 and B11 in \citetalias{stevens16} have $r_{\rm d}$ replaced by $r_{\rm s}$).  In a new feature (Appendix \ref{ssec:dispersion}), we also use this to inform the scale below which dispersion support is important in the disc.  Qualitatively consistent with the results presented in \citetalias{stevens16}, \DS~stellar discs have steeper-than-exponential profiles towards their centres (where pseudo-bulges can contribute to what is classed as the `disc' in the model).  This does not make extracting a meaningful radius (i.e.~$r_{\rm s}$) for aspects of the disc which \emph{are analytically modelled} to be exponential immediately straightforward.

What \emph{can} be directly measured for the discs are radii enclosing arbitrary fractions of their total mass.  While for exponential profiles there is a 1--1 relationship between any of these values (and the exponential scale radius), there can be significant scatter in the ratio of any two such radii for \DS~discs.  In an attempt to mediate this, we calculate the effective scale radius of the disc as
\begin{equation}
r_{\rm s} = \frac{0.596\, r_{*,50} + 0.257\, w_{90}\,  r_{*,90}}{1 + w_{90}}\,,
\label{eq:rs}
\end{equation}
where $r_{*,x}$ is the radius encompassing $x$ per cent of the stellar disc's mass (drawn directly from linear interpolation of the cumulative mass profile of the disc annuli), and $w_{90}$ is a free parameter that weights the average.  The coefficients in the numerator relate the corresponding $r_{*,x}$ values to the scale radius of an exponential disc.

The introduction of the $w_{90}$ parameter is important because $r_{*,50}$ probes the steepest part of the disc profile, whereas $r_{*,90}$ probes the region that can be approximated as exponential.  Higher values of $w_{90}$ will hence lead to larger values for $r_{\rm s}$.  Because the stellar dispersion profile assumes $r_{\rm s}$, $w_{90}$ sets the radius at which stellar discs are stable.  Because a stellar-driven instability in an annulus is resolved in \DS~by transferring stellar mass to adjacent annuli, where this can cascade through inner annuli all the way to the disc centre, the net effect is that discs shrink.  This is compounded by the instability-driven bulge also being smaller (in size, not mass), as this raises the circular velocity towards the disc centre, thereby lowering the radius for each annulus, which is fixed in terms of $j$.  A higher occurrence of disc instabilities leads to larger bulge masses.  As a result, $w_{90}$ controls not only the effective maximum mass a disc can have, but also the ratio of bulge- to disc-dominated systems (as a function of stellar mass).  We therefore chose a round value of $w_{90} \! = \! 2$, which approximately recovers the observed contributions of bulge- and disc-dominated galaxies to the stellar mass function, based on data from \citet{moffett16}.  This is shown and discussed further in Appendix \ref{app:constraints}.

\subsection{Dispersion support in discs}
\label{ssec:dispersion}
Previous versions of \DS~assumed that discs were in precise centripetal motion everywhere (i.e.~at every annulus boundary).  But, physically, one should expect that towards the centres of discs, random motions and pressure support (for gas) should be comparable to circular motion.  
To account for this, we now include an approximate consideration of dispersion support in discs in \DS.  Specifically, we modify the $j$-to-$r$ conversion for annuli by including a factor, $f_{\rm rot}(r)$, for the fraction of gravity balanced by rotational (circular) motion:
\begin{equation}
j = r\,v_{\rm circ}(r) = \sqrt{f_{\rm rot}(r)\, G\, M(<\!r)\, r}\,,
\label{eq:j2r}
\end{equation}
which is solved iteratively.  Here, $G$ is the gravitational constant, and $M(<\!r)$ is \emph{all} enclosed mass within $r$ \citepalias[for full details, see appendix B of][]{stevens16}.  The remaining fraction of gravity is assumed to be balanced by dispersion support and/or pressure.

We note that Equation (\ref{eq:j2r}) is based on a spherical gravitational potential.  In the inner parts of discs, the potential should instead be axisymmetric. Unfortunately, this calculation is already a bottleneck in the code, and so to employ a fully self-consistent consideration of the disc's potential would simply be too computationally demanding with model's current design.  The cost here is that we currently overestimate $v_{\rm circ}$ at low $r$, thereby underestimating $r$ for fixed $j$.  This means disc centres in the model are too dense, a problem identified in \citetalias{stevens16}.  The inclusion of $f_{\rm rot}$, where $f_{\rm rot}(r) \! \leq \! 1\, \forall\, r$, helps towards alleviating this, as it increases the corresponding $r$ for fixed $j$. Unfortunately, by itself, it is not enough to resolve this feature of the model.  A more self-consistent treatment of the disc potential will be incorporated in the future.  We do not expect the main scientific conclusions of this paper to be significantly affected by this.

The functional form of $f_{\rm rot}(r)$ should tend to zero at $r\!=\!0$ but rise to 1 over a short distance as dispersion support becomes negligible in a disc.  Recent analysis of multi-slit data capturing extended kinematics in lenticular galaxies (thought to be faded discs) by \citet{bellstedt17} provides an observational basis for this.  Those authors measured radial profiles for a \emph{local} stellar spin parameter, denoted $\lambda(R)$, which probes the ratio of rotation-to-(rotation+dispersion) support as a function of radius, at that radius.  Proxying $f_{\rm rot}(r)$ for $\lambda(R)$, and extrapolating their results for lenticular galaxies to be applicable to discs in general, implies that $f_{\rm rot}(r) \appropto 1 \! - \! e^{-r/r_0}$ (where $r_0$ is an as-yet-unspecified reference radius) broadly captures the behaviour we wish to model (cf.~the top panel of their figure 11).

Caution should be heeded in this extrapolation for a few reasons.  First, the sample size that this is based on is limited.  Second, multiple formation mechanisms for lenticular galaxies have been proposed \citep[cf.][]{lau06,cortesi11,quer15,bellstedt17}, so their kinematics might not be entirely representative of typical discs.  Third, and perhaps most importantly, these observations only pertain to $z\!\simeq\!0$.  At high redshift, galaxies are known to have lower values of $V/\sigma$, signifying that dispersion support plays a larger role \citep[see, e.g.,][and references therein]{wisnioski15,stott16}.  Coupled with this, the assumption that discs are thin (inherent to \DS) breaks down too.  We therefore focus on \DS~galaxies at $z\!\leq\!2$, although, more conservatively, our results are reliable at $z\!\lesssim\!1$.  While low-$z$ galaxies descend from high-$z$ galaxies where model assumptions about discs are inaccurate, provided there has been sustained low-$z$ disc growth that exceeds any inherited high-$z$ growth, the inherited high-$z$ uncertainties are diminished.  As shown by \citet{lagos17}, the formation of \zo~discs happened predominantly at $z\!<\!1.5$.  Currently, we are unaware of other published results from observational data that directly measure a quantity that resembles $f_{\rm rot}(r)$ for galaxy discs.

In addition to the above, one might expect that a disc embedded in a galaxy with a large bulge will find dispersion support to be more relevant.  Indeed, should this be the case, the inner parts of discs would become less dense and hence less susceptible to star formation.  This would qualitatively account for the phenomenon of morphological quenching \citep{martig09}.  With all the above in mind, we implement
\begin{equation}
f_{\rm rot}(r) = 1 - \exp\left[-\frac{3\, r}{r_s} \left(1 - \frac{m_{\rm *,bulge}}{m_*}\right) \right]\,,
\label{eq:frot}
\end{equation}
where $r_s$ is from Equation (\ref{eq:rs}).  We note that while Equation (\ref{eq:j2r}) is calculated once per annulus per galaxy per sub-time-step in the model,%
\footnote{In \DS~there are 10 `sub-time-steps' between each snapshot fed in from the merger trees, wherein all the galaxy evolution physics is calculated.} 
$r_{\rm s}$ is updated not only when this happens, but also whenever star formation or instabilities occur.  The factor of 3 within the square brackets of Equation (\ref{eq:frot}) approximately recovers where the $\lambda(R)$ profiles in the results of \citet{bellstedt17} asymptote for most lenticular galaxies.  The precise form of Equation (\ref{eq:frot}) has room to be modified as new data are published.

We have also relaxed the previously imposed limit on rotation curves that they could not exceed the halo $V_{\rm max}$ in the merger trees.  With dispersion support now considered, fewer galaxies' rotation velocities saturate at low radius.  The $V_{\rm max}$ of a halo should be affected by baryonic physics \citep[as per the effects of baryons on total density profiles: see, e.g.,][]{dicintio14,brook15,schaller15a,peirani17,remus17,bellstedt18}, and hence galaxies should be allowed to rotate faster than implied by a pure $N$-body simulation.  We still maintain a limit on rotation velocities, but now set this (somewhat arbitrarily) to twice the halo $V_{\rm max}$.


\section{Observational constraints}
\label{app:constraints}

We summarise here the observational constraints used for calibrating \DS, and present all calibration figures.  While the calibration was performed using the Mini-Millennium simulation (box length 62.5$h^{-1}$\,cMpc), these figures use the full simulation (box length 500$h^{-1}$\,cMpc).  In each figure, we have added the most recent observational data as a point of comparison.  These updated data were not used during the calibration procedure.  Given the leniency imposed by \DS~being calibrated manually, had we calibrated on these data, our final parameters likely would not have changed.  Note that to highlight the resolution limit of the simulation+model, in all following figures, we include dashed and solid curves pertaining to all haloes with at least 20 particles ($N_{\rm p}\!\geq\!20$), and all haloes with at least 100 particles at some point in their history ($N_{\rm p,max}\!\geq\!100$), respectively.  The minimum mass shown in the plots (the left bound of the axes in most cases) is the median value for galaxies with $N_{\rm p,max}\!\geq\!100$ (galaxies with no mass of the relevant quantity were excluded for calculating this value).

\begin{figure*}
\centering
\includegraphics[width=\textwidth]{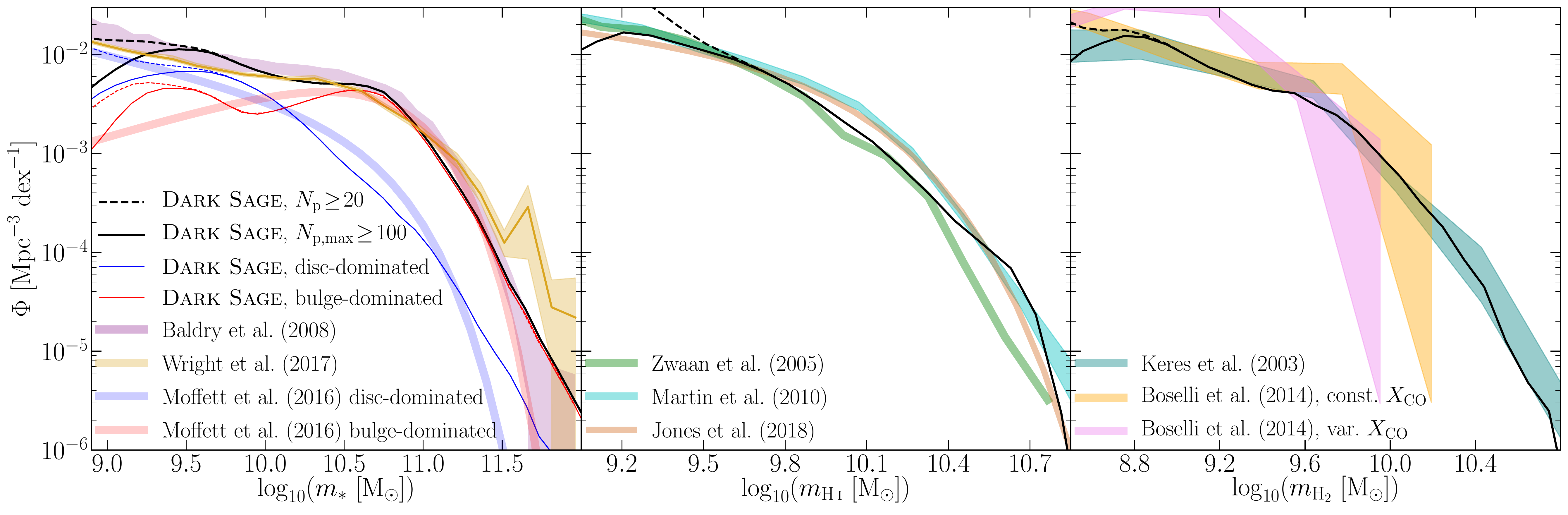}
\caption{Respective mass functions for the stellar, \HI, and \Htwo content of \DS~galaxies at $z\!=\!0$, compared against the observational data used to constrain the model's free parameters, and the latest observational data from other sources.  $N_{\rm p}$ denotes number of particles in the associated haloes, whereas $N_{\rm p,max}$ denotes the historical maximum number of particles of those haloes.}
\label{fig:mfs}
\end{figure*}

In Fig.~\ref{fig:mfs} we present the stellar, \HI, and \Htwo mass functions for \DS~at $z\!=\!0$.  The stellar mass function was calibrated to \citet{baldry08}, as were the bulge- and disc-dominated stellar mass functions to \citet{moffett16}.  The \citet{baldry08} data are as presented in their paper, whereas we use the Schechter function fits from \citet{moffett16}.  For the latter, `bulge-dominated' galaxies are those of Hubble type E/S0-Sa, and `disc-dominated' are Sab-Scd/Sd-Irr.  Because these are qualitative, visually measured classifications, there is uncertainty associated with comparing to quantitatively distinct bulge- and disc-dominated galaxies in \DS.  For this reason, we did not over-fit the $w_{90}$ parameter that solely controlled the ratio of bulge- to disc-dominated galaxies, opting for a round value of 2 that sufficiently represented the observations.  Also compared in Fig.~\ref{fig:mfs} is the most recent stellar mass function from the GAMA survey \citep{wright17}.  

The \HI~mass function was calibrated using data from \citet{zwaan05} and \citet{martin10}.  The latter comes from the ALFALFA survey, which has since been updated in \citet{jones18}; the Schechter function fit to their data is included in Fig.~\ref{fig:mfs}. The \Htwo mass function was calibrated against \citet{keres03}, who assumed a constant $X_{\rm CO}$ factor in converting CO luminosity to \Htwo mass.  Compared are data from \citet{boselli14b}, using both constant and variable (i.e.~luminosity-dependent) $X_{\rm CO}$ factors.

\begin{figure}
\centering
\includegraphics[width=\textwidth]{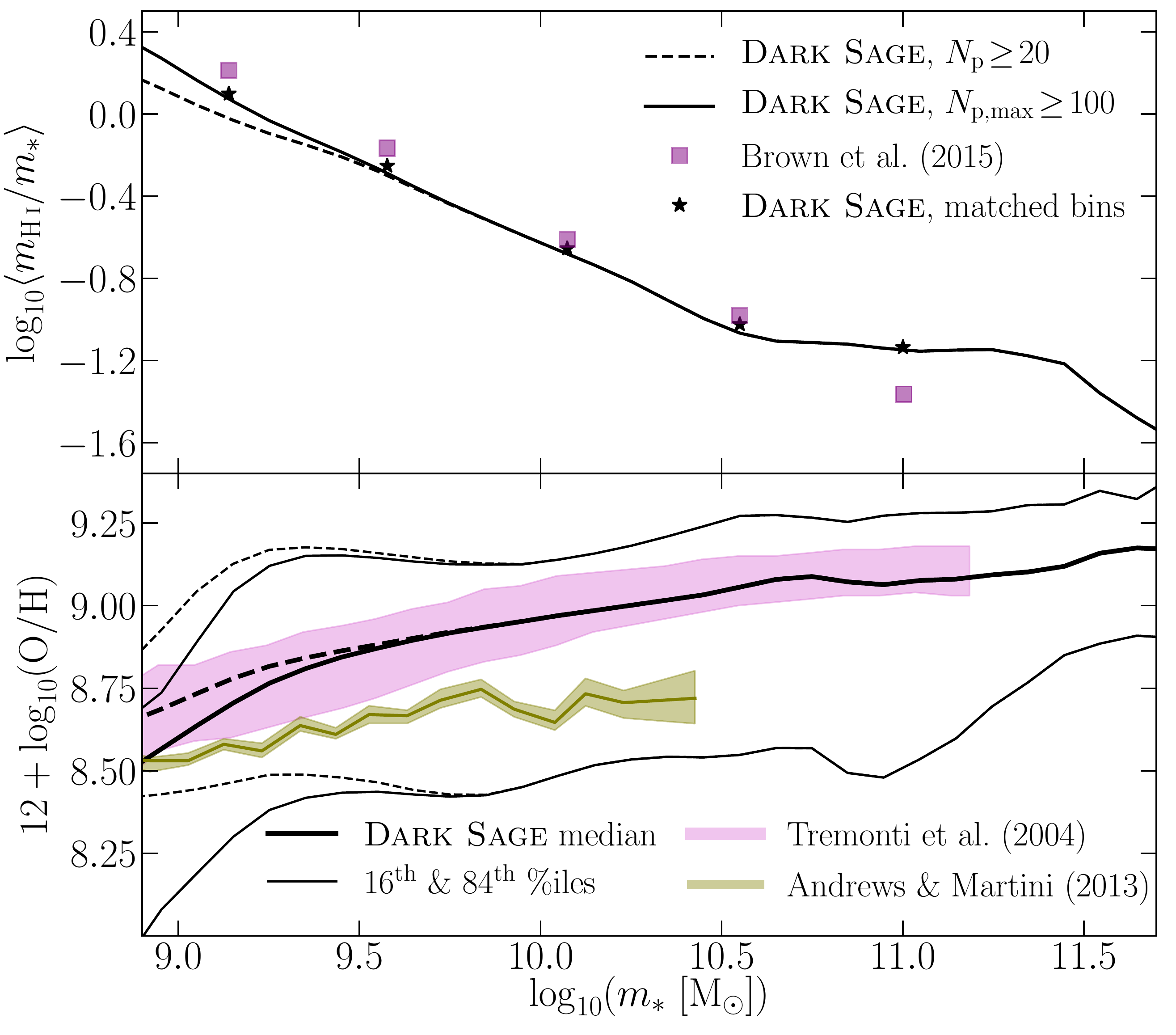}
\caption{\HI~fraction (top) and gas metallicity (bottom) as a function of stellar mass for \DS~galaxies at $z\!=\!0$.  The \HI~fraction shows the mean for both the model and constraint data, where the `matched bins' were used for calibration.  The shaded region for \citet{tremonti04} covers the 16$^{\rm th}$--84$^{\rm th}$ percentiles, while the uncertainty on the mean relation is shown for \citet{andrews13}.}
\label{fig:mstarx}
\end{figure}

Both the mass--metallicity relation and mean \HI~fraction as a function of stellar mass are presented in Fig.~\ref{fig:mstarx}.  For the \HI~fractions, the model was calibrated against the stacked mean relation from ALFALFA data \citep{brown15}.  In \citetalias{sb17}, this was the strongest constraint for \DS, with a relatively precise fit.  Here, the stringency of this has been loosened.  The median relation for the mass--metallicity relation was calibrated against that from \citet{tremonti04}.  For \DS, we took $12\!+\!\log_{10}({\rm O/H}) \!=\! 9\!+\!\log_{10}(Z/0.02)$.  While the median \DS~relation agrees well with \citet{tremonti04}, the scatter in \DS~is far larger (this was also true for \citetalias{stevens16} -- cf.~their fig.~A6).  Investigation into the cause of this is beyond the scope of this paper, but is likely to prove informative for future developments of the model.  We have also compared the same relation from \citet{andrews13}, whose results have galaxies as systematically less metal-rich.  Where the \citet{tremonti04} metallicities are based on strong-line diagnostics, those of \citet{andrews13} include a direct measurement of the electron temperature after stacking spectra (and hence the scatter is not probed -- instead only the error on the mean is given).

\begin{figure}
\centering
\includegraphics[width=\textwidth]{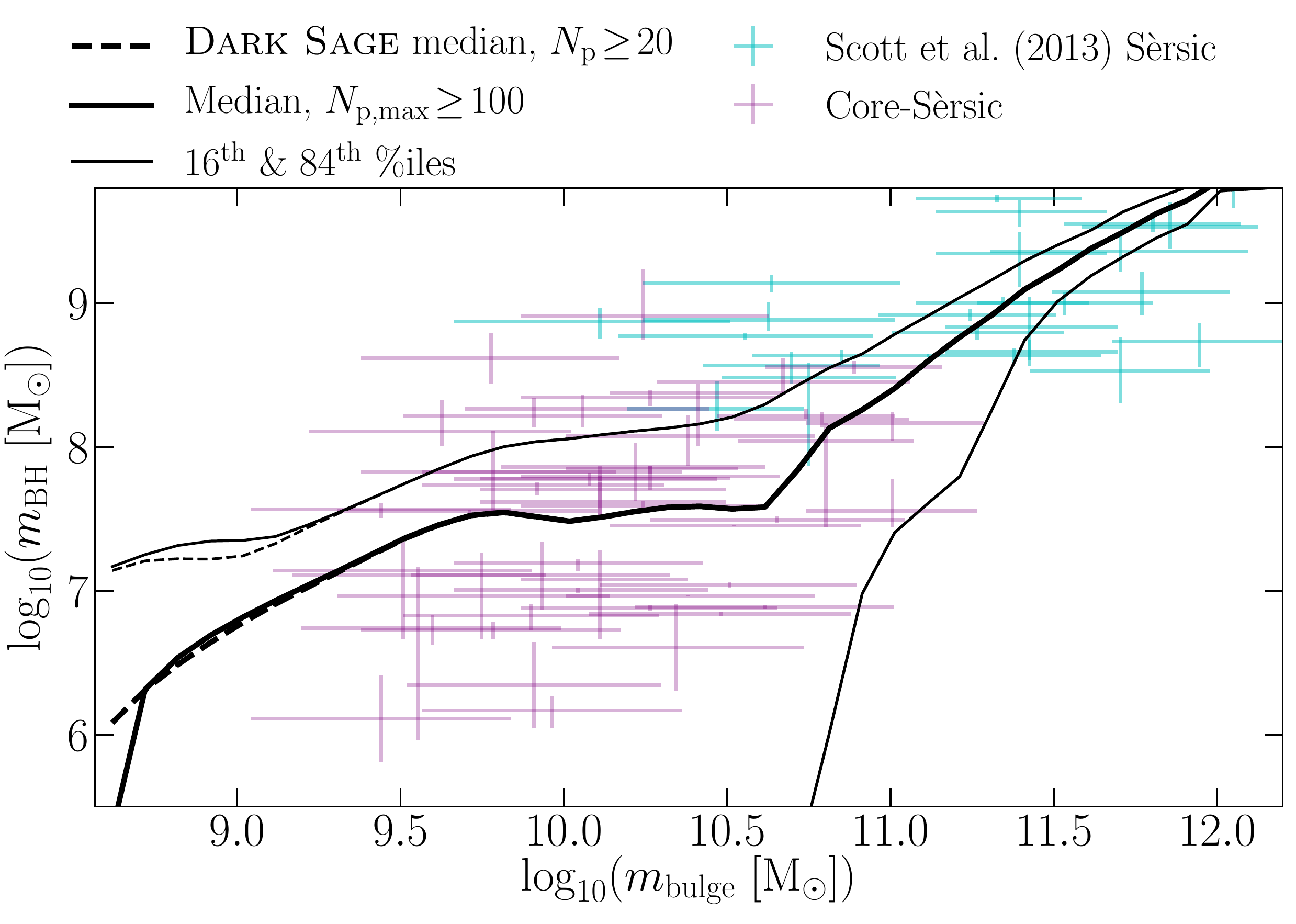}
\caption{Black hole--bulge mass relation for $z\!=\!0$ \DS~galaxies, as calibrated to the observational data from \citet{scott13}.}
\label{fig:bhbm}
\end{figure}

The black hole--bulge mass relation is presented in Fig.~\ref{fig:bhbm}.  This was calibrated against data from \citet{scott13}.  While those authors split galaxies into S\`{e}rsic and core-S\`{e}rsic types, this distinction was not considered for calibrating \DS.  We simply aimed for a median relation that met the data, with a comparable level of scatter.  Both instability- and merger-driven bulges contribute to \DS~results in Fig.~\ref{fig:bhbm}, but pseudo-bulges do not (as per their being inherently part of the disc).

\begin{figure}
\centering
\includegraphics[width=\textwidth]{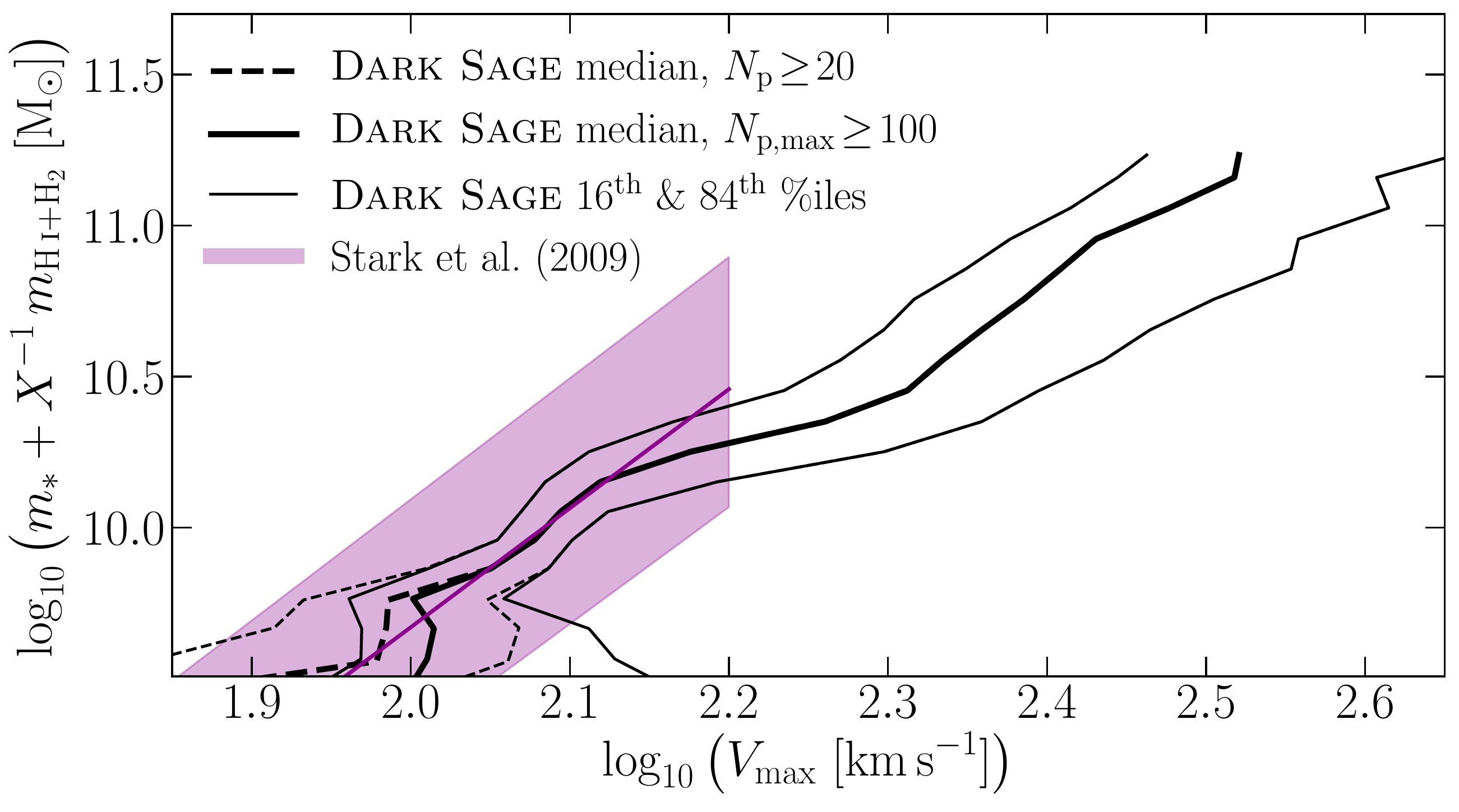}
\caption{Baryonic Tully--Fisher relation for gas-dominated \DS~galaxies with stable rotation curves.  The shaded region of \citet{stark09} covers their best-fitting power law, including the random uncertainties associated with that fit (but excluding systematics), approximately out to where their data apply.  Running medians and percentiles for \DS~are calculated in bins along the $y$-axis (whereas all previous plots were done along the $x$-axis).}
\label{fig:btf}
\end{figure}

In Fig.~\ref{fig:btf}, we show the Baryonic Tully--Fisher relation for \DS~galaxies.  These were calibrated to meet the power law fit from \citet{stark09}.  To match the observed sample, we only selected \DS~galaxies that were  gas-dominated (i.e. where $X^{-1}m_{\rm H\,{\LARGE{\textsc i}}+H_2} \! > \! m_*$) and whose rotation velocities varied by either less than 15\% between $3\,r_{\rm s}$ and the outermost annulus or less than 10\% between $4\,r_{\rm s}$ and the outermost annulus.  While in \citetalias{stevens16} virial velocity was used in place of maximum velocity, we now actually use the peak velocity from the rotation curves that \DS~produces whenever $V_{\rm max}$ is concerned.

\begin{figure}
\centering
\includegraphics[width=\textwidth]{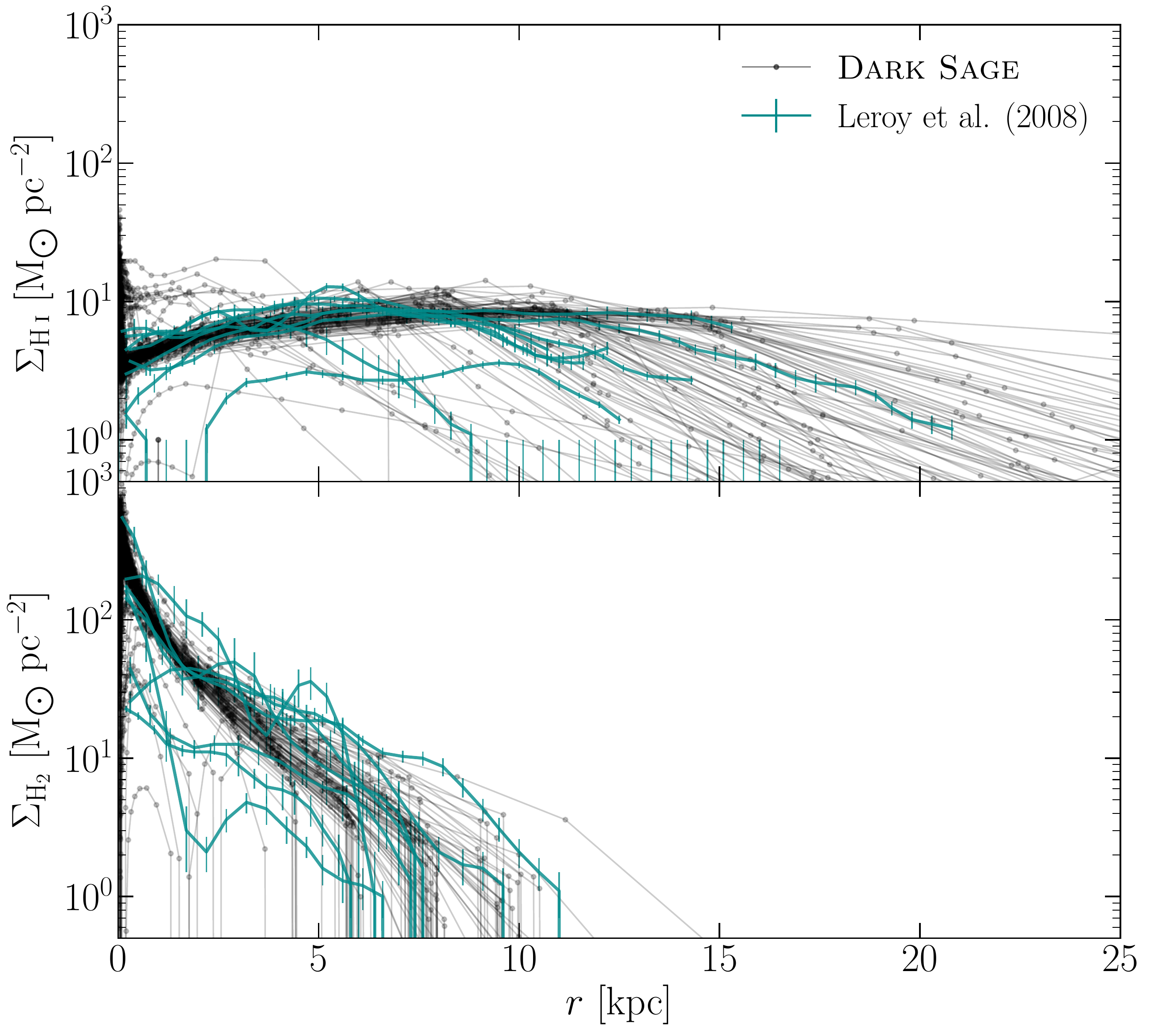}
\caption{Atomic- and molecular-hydrogen surface density profiles for \DS~galaxies that are of comparable mass and morphology to the compared galaxies from \citet{leroy08}, used in calibrating the model.  We show profiles for 100 random \DS~galaxies that make the cut.}
\label{fig:profs}
\end{figure}

Finally, in Fig.~\ref{fig:profs}, we present \HI~and \Htwo surface density profiles for a sample of galaxies that are directly comparable to those from \citet{leroy08}.  The selected \DS~galaxies are disc-dominated centrals with $175 \! \leq \! V_{\rm max} / ({\rm km\,s}^{-1}) \! \leq \! 235$, $m_* \! \geq \! 10^{10}\,{\rm M}_{\odot}$, and $m_{\rm gas} \! \geq \! 10^{9.2}\,{\rm M}_{\odot}$.  Because we have now calibrated on these profiles, it is no surprise we get a better match than in \citetalias{stevens16}.

We note that we abandoned the cosmic star formation density history of the Universe as a constraint entirely.  Part of the reason why \DS~struggles to reproduce the observed relation is in its assumption that gas in a galaxy sits in a thin, predominantly rotationally supported disc; this assumption breaks down at high redshift (as noted in Appendix \ref{ssec:dispersion}).  But other semi-analytic models still make this assumption at some level and are able to recover reasonable cosmic star formation histories.  We have found that changing the prescription for the mass-loading from stellar feedback from one that scales with local surface density to, e.g., one that is uniform has a significant impact on this relation (and many others).  In future work, we intend to revamp stellar feedback in \DS~and reassess the impact this has on the cosmic star formation history.

\end{document}